\documentclass[aps,pra,twocolumn,groupedaddress,floatfix]{revtex4-1}

\usepackage{graphicx}
\usepackage{dcolumn}
\usepackage{bm}
\usepackage{amsmath}

\begin{document}

\title{Bottom-up configuration-interaction emulations of ultracold fermions in entangled optical plaquettes: building blocks of unconventional superconductivity}

\author{Benedikt B. Brandt}
\email{benbra@gatech.edu}
\author{Constantine Yannouleas}
\email{Constantine.Yannouleas@physics.gatech.edu}
\author{Uzi Landman}
\email{Uzi.Landman@physics.gatech.edu}

\affiliation{School of Physics, Georgia Institute of Technology,
             Atlanta, Georgia 30332-0430}

\date{17 March 2017}

\begin{abstract}
A microscopic configuration-interaction (CI) methodology is introduced to enable bottom-up 
Schr\"{o}dinger-equation emulation of unconventional superconductivity in ultracold optical traps. 
We illustrate the method by exploring the properties of $^6$Li atoms in a single square plaquette 
in the hole-pairing regime, and by analyzing the entanglement (symmetry-preserving) and disentanglement 
physics (via symmetry-breaking, associated with the separation 
of charge and spin density waves) of two coupled plaquettes in the same regime. The single-occupancy 
RVB states contribute only partially to the exact many-body solutions, and the CI results map onto a 
Hubbard Hamiltonian, but not onto the double-occupancy-excluding $t$-$J$ one.
For the double-plaquette case, effects brought about by breaking the symmetry between two weakly-interacting  
plaquettes, either by distorting, or by tilting and detuning, one of the plaquettes with respect to 
the other, as well as spectral changes caused by increased coupling between the two plaquettes, are explored.
\end{abstract}

\maketitle

\section{Introduction}

Rapid experimental advances in the creation of finite systems of ultracold atoms 
using few optical traps (bottom-up) \cite{joch12,joch15,joch15.2,kauf14,kauf15} or through the use of 
extended optical lattices (top-down) \cite{hule15,bloc16,grei16,zwie16} are promising approaches toward 
realization of Feynman's vision of a quantum simulator \cite{feyn82}, capable of finding solutions to systems 
that are otherwise numerically and/or analytically intractable. However, apart from a few double-well 
investigations \cite{yann15,yann16,poll15}, to date model-independent microscopic studies of multiwell systems 
providing theoretical insights and guidance to experimental efforts are largely lacking.

Here we introduce a configuration-interaction (CI) Schr\"{o}dinger-equation methodology 
\cite{yann07,yann09,yann15,yann16,szabobook} for exploring finite plaquette systems
assembled from individual optical traps; for a brief description of the CI method, see Appendix \ref{cime}. 

These systems are fundamental stepping stones toward bottom-up
realization of large scale checkerboard or square ulracold atom lattices which are promising candidates for 
emulating the physics underlying $d$-wave high-T$_c$ superconductivity \cite{ande08,fuku08,scal96,tsai06,tsai08} 
in optical lattices \cite{hofs02,zoll06,rey09,pare08,nasc12,hule15,bloc16,grei16,zwie16}. The work described here, 
demonstrating the feasibility of such exact CI calculations for interatomic contact interactions, can be extended
to electronic plaquettes, i.e., to quantum-dot-arrays, governed by long-range Coulomb interactions.

The plan of the paper is as follows:

In section \ref{sp} we explore first  the properties of ultracold fermionic atoms ($^6$Li) confined in a single 
square plaquette (4-sites) in the regime of hole pairing, and subsequently analyze the entanglement physics
of two coupled plaquettes in the hole-pairing regime. 

In section \ref{dp}, for the “hole-doped” coupled plaquettes (8-sites, 
six atoms), we analyze the wave function anatomy of the entangled \cite{yann15,yann16,poll15}
(Schr\"{o}dinger-cat) two almost-degenerate manifolds, comprising: (${\it A}$) the ground state (GS) and 1st excited 
(1EX) state, and (${\it B}$) the two higher excited states (2EX and 3EX). The almost-degenerate states  have good, but 
opposite, parities. When symmetry-broken (SB) -- either by superposing $(\pm)$ the degenerate pair in each 
manifold, or via  offsetting the energies (tilting) of the two 
plaquettes -- the SB states coming from ${\it A}$ are characterized by a particle (``charge'')-density 
modulation (``wave''), i.e, a CDW portraying the hole-paired, (4,2)
or (2,4), components, whereas the SB states originating from ${\it B}$ remain in a (3,3) particle distribution
(each of the plaquettes having an unpaired hole), but exhibit a spin-polarizarion density modulation (``wave''),
i.e., a SDW. For the the double-plaquette case, we further explore effects brought about by breaking the symmetry 
between two weakly-coupled plaquettes, either by distorting one of the plaquettes, or by tilting and detuning  
one of the plaquettes with respect to the other. Spectral changes caused by increased tunnel coupling (e.g., by 
decreased inter-plaquette distance) between the two plaquettes are also considered.

Noteworthy is our finding that the GS $d$-wave resonating valence bond (RVB) state contributes only partially 
to the exact many-body wave function -- i.e.,  double occupancies (referred also as doublons) need be included. 
Indeed, our microscopic results map properly onto a Hubbard Hamiltonian (including extended Hubbard models 
\cite{fali69,lueh15} depending upon the range of the experimental parameters), but not onto the 
double-occupancy-excluding $t$-$J$ model \cite{ande08,rice88,fuku08}. Our conclusions agree with recent 
\cite{bloc16,grei16,zwie16} observations of doublons in two-dimensional (2D) optical lattices.

The rest of the paper comprises the Summary (Sec.\ \ref{summ}) and three Appendices, concerning 
a brief description of the CI method (Appendix \ref{cime}), the mathematical definitions of single-particle
densities and two-body and $N$-body conditional probability distributions (Appendix \ref{defs}), and the
specification of the correspondence between spin and RVB functions and CI many-body wave functions
(Appendix \ref{spin}).  

\section{The single plaquette}
\label{sp}

The short-range 2-body repulsion in the Hamiltonian is described by a Gaussian of width $\sigma$, i.e., by
\begin{equation} 
V({\bf r}_i,{\bf r}_j)=\frac{\lambda}{\sigma^2\pi}e^{-({\bf r}_i-{\bf r}_j)^2/\sigma^2}.
\label{2bi}
\end{equation}
$\lambda$ is the strength parameter. Here and throughout the paper: $\sigma = \sqrt{2} l_0/10 = 0.1833$ $\mu$m, 
where the oscillator length $l_0^2 = \hbar/(M_{^6{\rm Li}}\omega$), $M_{^6{\rm Li}}$ being the $^6$Li mass and
$\hbar\omega  = 1$ kHz being the trapping frequency of the plaquette potential wells \cite{note1}.
This form of interaction provides a good approximation for the atom-atom 
interactions and avoids the peculiarities of the delta function in two dimensions \cite{doga13}.

\begin{figure}[t]
\centering\includegraphics[width=7cm]{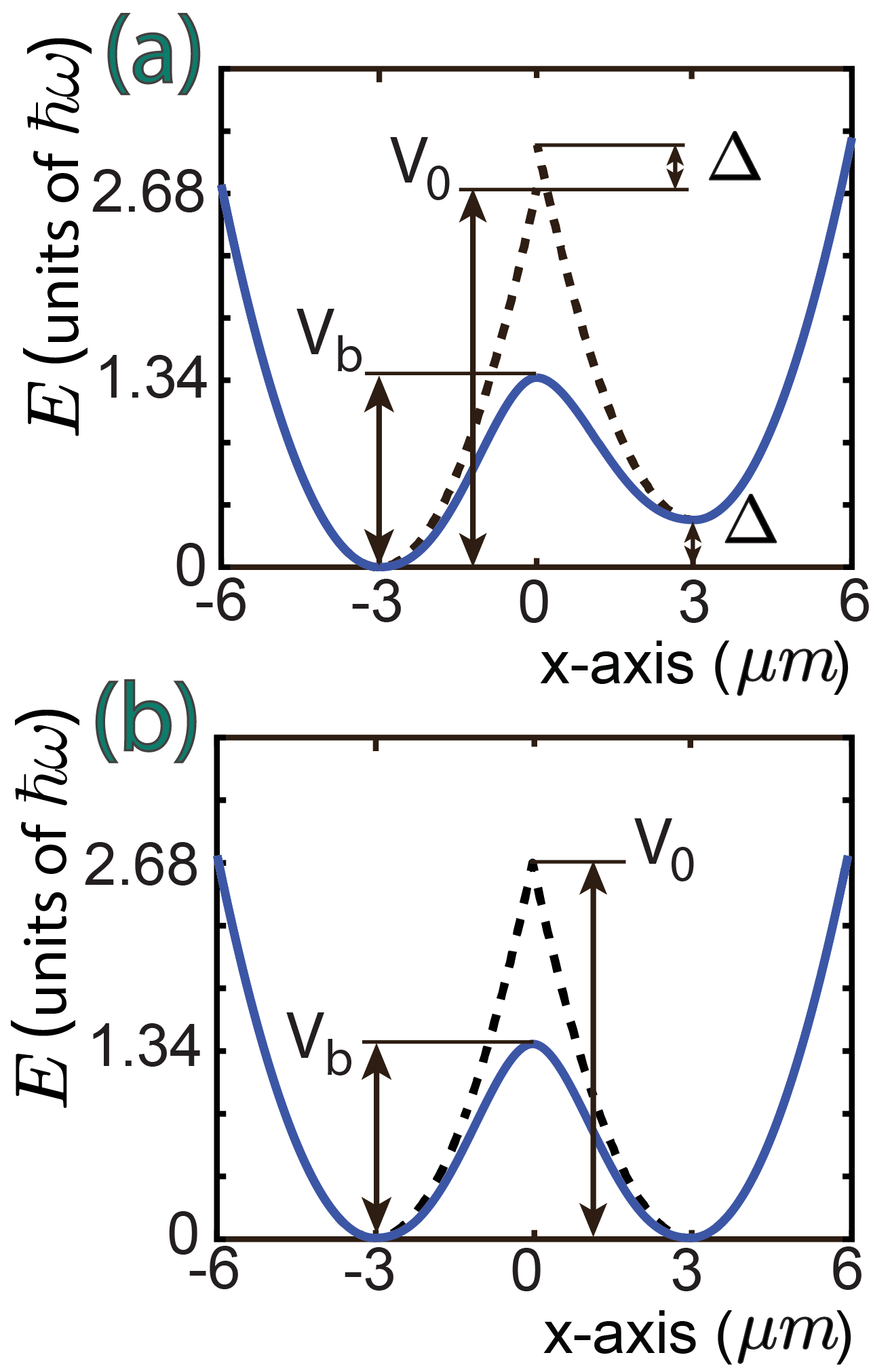}
\caption{
Illustration of the one-dimensional TCO potential with a smooth neck
(solid lines). (a) Case with a tilt $\Delta>0$. (b) Case without a tilt. $\hbar \omega$ is the
trapping frequency (given in energy units of kHz in this paper) of the bare harmonic oscillator
(dashed lines) for each site. 
}
\label{fig1_t}
\end{figure}

The potential surface of the 4-site (or 8-site) plaquette is constructed with the help of the 
two-center/smooth-neck oscillator Hamiltonian which was previously introduced in Refs.\ 
\cite{yann09,yann15,yann16,yann99}. 
The smooth-neck interwell barrier $V_b$ can be varied independently and is controlled by the parameter 
$\epsilon_b=V_b/V_0$, where $V_0$ is the intersection height of the two bare potential parabolas from neighboring sites,
i.e., prior to inserting the smooth-neck potential contribution \cite{yann15,yann09,yann99}. For a graphical 
illustration of the two-center-oscillator (TCO) double-well potential, 
including the definitions of $V_b$ and $V_0$, see Fig.\ \ref{fig1_t}. Here and throughout the paper: the intersite 
distance in a single plaquette is $d_w = 6$ $\mu$m and $\epsilon_b=0.5$ 
(yielding $V_b=1.34$ kHz), unless noted otherwise. 
The potential surfaces of the 4-site and 8-site plaquettes is constructed by combining such TCO potentials along the $x$ 
and $y$ directions (see Supplemental Material for details \cite{suppl}).

\begin{figure}[t]
\centering\includegraphics[width=8.0cm]{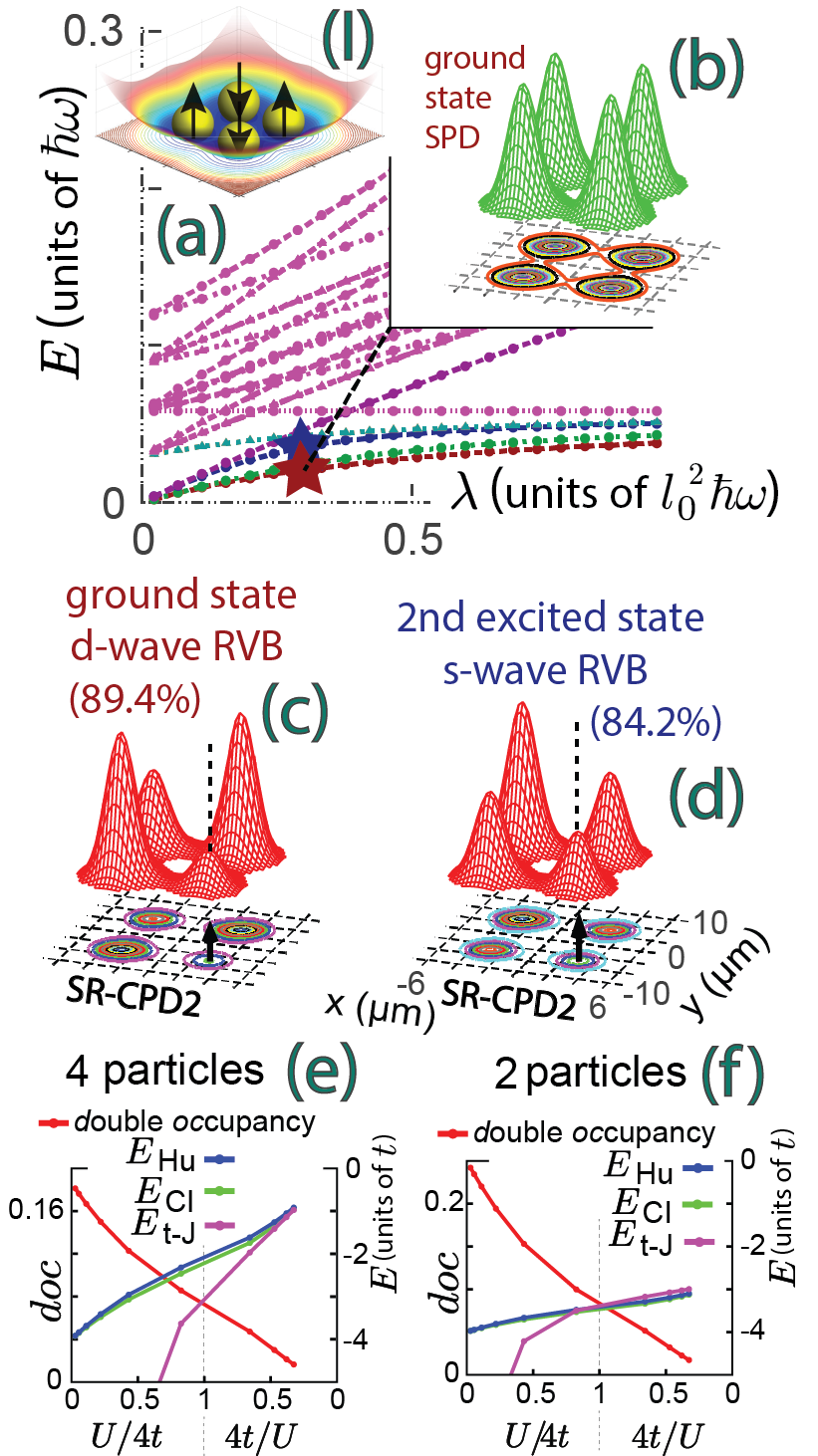}
\caption{
CI results for $N=4$ (a,b,c,e) and $N=2$ (d,f) $^6$Li atoms in a 4-site plaquette. (a)
Energy spectrum. $V_b=1.34$ kHz.  (b) GS SPD [red star in (a)]. (c,d) SR-CPD2s for GS and 2EX.
Up black arrows: up-spin at the observation point. Red humps: probability for down spin atoms. 
Note the double-occupancy hump at the observation point. (e,f) GS total energies ($E$, in units of $t$) 
for the CI, Hubbard, and $t$-$J$ Hamiltonians; CI double-occupancy (doc) in red. 
Inset (I) illustrates the 4-site external potential.
}
\label{fig2_t}
\end{figure}

The CI single-plaquette low-energy spectrum as a function of the two-body repulsion strength $\lambda$,
is displayed in Fig.\ \ref{fig2_t}(a), with the single-particle density (SPD) for the GS (see red star), or 
for the second excited state (blue star), being shown in Fig.\ \ref{fig2_t}(b); for the definition of the SPD
in the CI approach, see Appendix \ref{defs}. A total-spin projection $S_z=0$ has been assumed. 
The CI wave functions depend both on the atom (continuous) position, ${\bf r}_i$, and the spin, $\sigma_i$, 
$i=1,2,\ldots$, variables. However, for plaquettes with sufficiently high interwell barriers, 
the microscopic structure of a CI state $|\Phi\rangle$ can be mapped onto the usual Hubbard Hilbert space 
involving the superposition of many primitive basis functions 
$\Omega_i \equiv |\ldots d \ldots \uparrow \ldots \downarrow \dots 0 \ldots\rangle$; $d$ here denotes a doubly
occupied site, $\uparrow$ and $\downarrow$ denote the site's spin occupancy, and 0 denotes an empty site. 
Inset (I) in Fig.\ \ref{fig2_t} describes the 4-site external potential.
The coefficients $c_i$ of the primitives $\Omega_i$ 
in the expansion of $|\Phi\rangle$, can be extracted from the CI wave functions with the help of
the spin-resolved conditional probability distributions (SR-CPDs) \cite{yann15,yann16,yann07.2}. 
Two SR-CPDs are used here: one (SR-CPD2), expressed as an expectation value over the many-body wavefunction,
describes the space and spin correlation of a particle pair, for a given location and spin
of one of them (referred to as an observation point) and the other being any other particle of the $N-1$ 
remaining ones (with a specified spin). The other probability distribution (SR-CPDN), given by the 
modulus square of the many-body function, expresses the spatial probability of finding the $N$-th particle with a 
specified spin when one fixes the positions and spins of the other $N-1$ particles (for the mathematical 
definition of SR-CPDs, see Appendix \ref{defs}). 

\begin{figure}[t]
\centering\includegraphics[width=7.4cm]{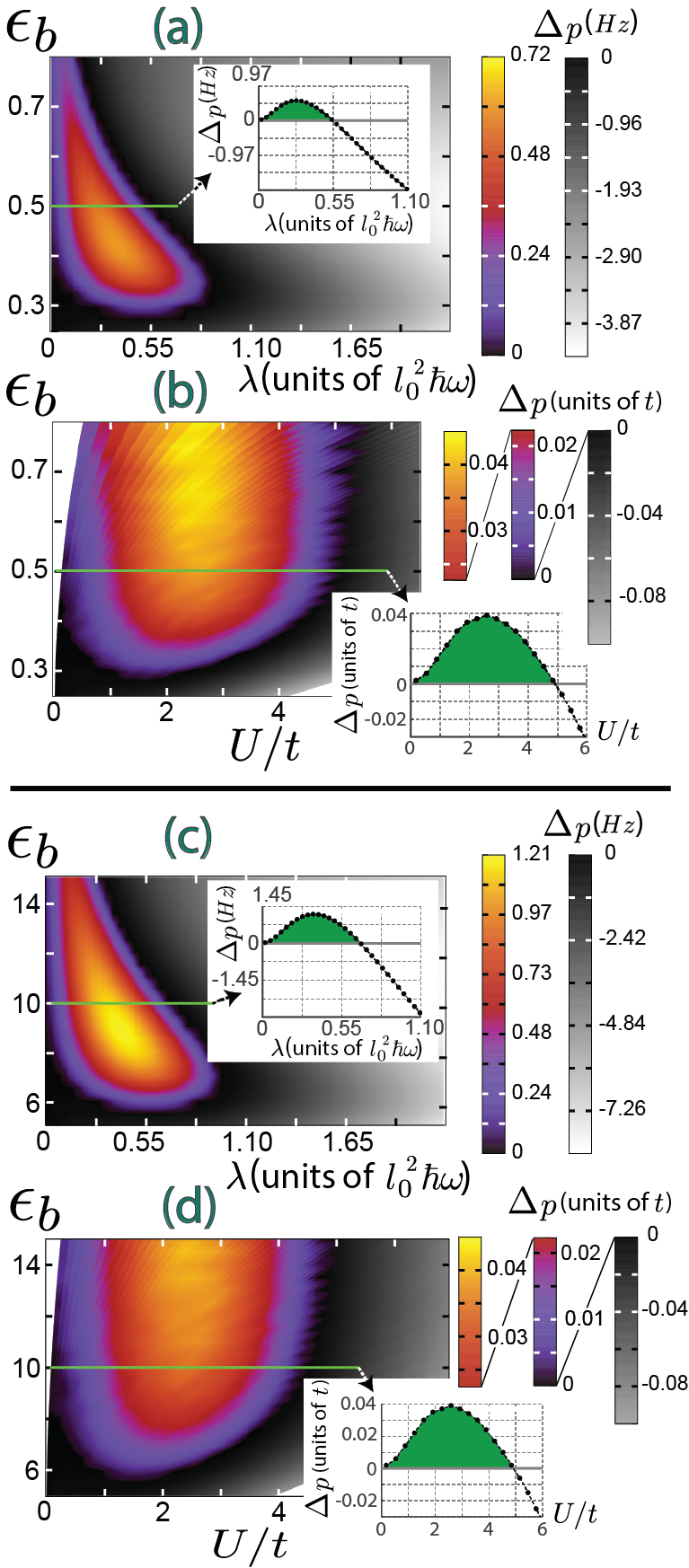}
\caption{
CI-calculated contour plot for $\Delta_p$ [Eq.\ (\ref{dgap})] for a single 4-well plaquette with 
$d_w=6$ $\mu$m (a,b) and $d_w=2.5$ $\mu$m (c,d) and variable $\epsilon_b$. Positive values (blue to yellow) 
indicate pair-binding. In (a,c) the horizontal axis is the repulsion strength $\lambda$, and in (b,d) 
the Extended-Hubbard parameter ratio $U/t$. In (a,c) $\Delta_p$ is in units of Hz, and in (b,d) in units of the 
corresponding hopping parameter $t$ between nearest-neighbor wells.
Parameters used in the calculations for the ground states of $N = 4$, $N = 3$, and $N = 2$ fermions are:
$\hbar\omega = 1$ kHz, $l_0=1.296$ $\mu$m, $\sigma=0.184$ $\mu$m, and $S=0$, $S_z=0$. $V_b$ varies in 
the ranges from $0.67$ kHz ($\epsilon_b=0.25$) to $2.14$ kHz ($\epsilon_b=0.8$) in (a,b), and 
$2.32$ kHz ($\epsilon_b=5$) to $6.98$ kHz ($\epsilon_b=15$) in (c,d). 
Insets show cuts for an intersite barrier $\epsilon_b = 0.5$ in (a,b) and $\epsilon_b = 10$ in (c,d). 
}
\label{fig3_t}
\end{figure}

Figs.\ \ref{fig2_t}(c,d) display the SR-CPD2s of a 4-fermion plaquette for the CI states (GS and 2EX) 
associated with the two stars in Fig.\ \ref{fig2_t}(a). The GS SR-CPD2 in Fig.\ \ref{fig2_t}(c) reveals the 
presence (89.4\%) of a $d$-wave RVB component, and the 2EX SR-CPD2 in Fig.\ \ref{fig2_t}(d) reveals the presence 
(84.2\%) of an $s$-wave RVB component (see the analysis in Appendix \ref{spin});
these percentages are derived from the double-occupancy fraction at the fixed-point site [i.e., the hump 
volume above the arrows in Figs.\ \ref{fig2_t}(c,d)]. 

The above SR-CPD2 analysis points to a deficiency of the $t$-$J$ model because it excludes \cite{triv04}
double occupancies. We further illustrate the limitations of the $t$-$J$ model for the 
full range of the repulsive interaction by plotting for all three approaches (CI, Hubbard, $t$-$J$) the 
4-fermion and 2-fermion total energies ($E$) for a single plaquette, as well as the double occupancy ($doc$),
as a function of $U/t$; see Figs.\ \ref{fig2_t}(e) and \ref{fig2_t}(f), respectively.
Evidently, the CI and Hubbard energies, apart from a constant shift, are in very good overall agreement, whereas 
the $t$-$J$ values deviate greatly. The exact CI results provide a here-to-date lacking benchmark for assessments
of the validity of the $t$-$J$ model and its variants \cite{ande08,rice88,fuku08,spal17}, as well as the 
Hubbard-calculated double occupancies.

The hole-pairing gap may be defined in terms of the energies of a 4-site plaquette, as follows \cite{tsai06}:
\begin{equation}
\Delta_p=2E_{\rm GS} (N=3)- [E_{\rm GS}(N=4) + E_{\rm GS}(N=2)],
\label{dgap} 
\end{equation}
where $N$ is the total number of fermions on the plaquette.

CI-calculated pairing gaps are shown in Fig.\ \ref{fig3_t} for two interwell distances: $d= 6$ $\mu$m 
[in (a,b)], and $d=2.5$ $\mu$m [in (c,d)]. 
Fig.\ \ref{fig3_t}(a) displays $\Delta_p$ as a function of the repulsion sterngth, $\lambda$, 
and the interwell parameter $\epsilon_b$ in a single plaquette. The gap maximum 
($\sim$ 0.72 Hz) occurs at ($\lambda \sim 0.30$ $l_{0}^2 \hbar \omega$, $\epsilon_b \sim 0.43$). 

To compare with the results of the Hubbard model, Fig.\ \ref{fig3_t}(b) displays the same CI results for 
$\Delta_p$, but with all energies expressed in units of the intersite tunneling parameter $t$ and the 
interaction strength $\lambda$ expressed as the ratio of $U/t$, with $U$ being the Hubbard on-site repulsion 
extracted from the CI calculation. The maximum of $\Delta_p$ is now $0.045t$, occurring 
at ($U/t=2.6$, $\epsilon_b \geq 0.7$).

\begin{figure}[t]
\centering\includegraphics[width=7.8cm]{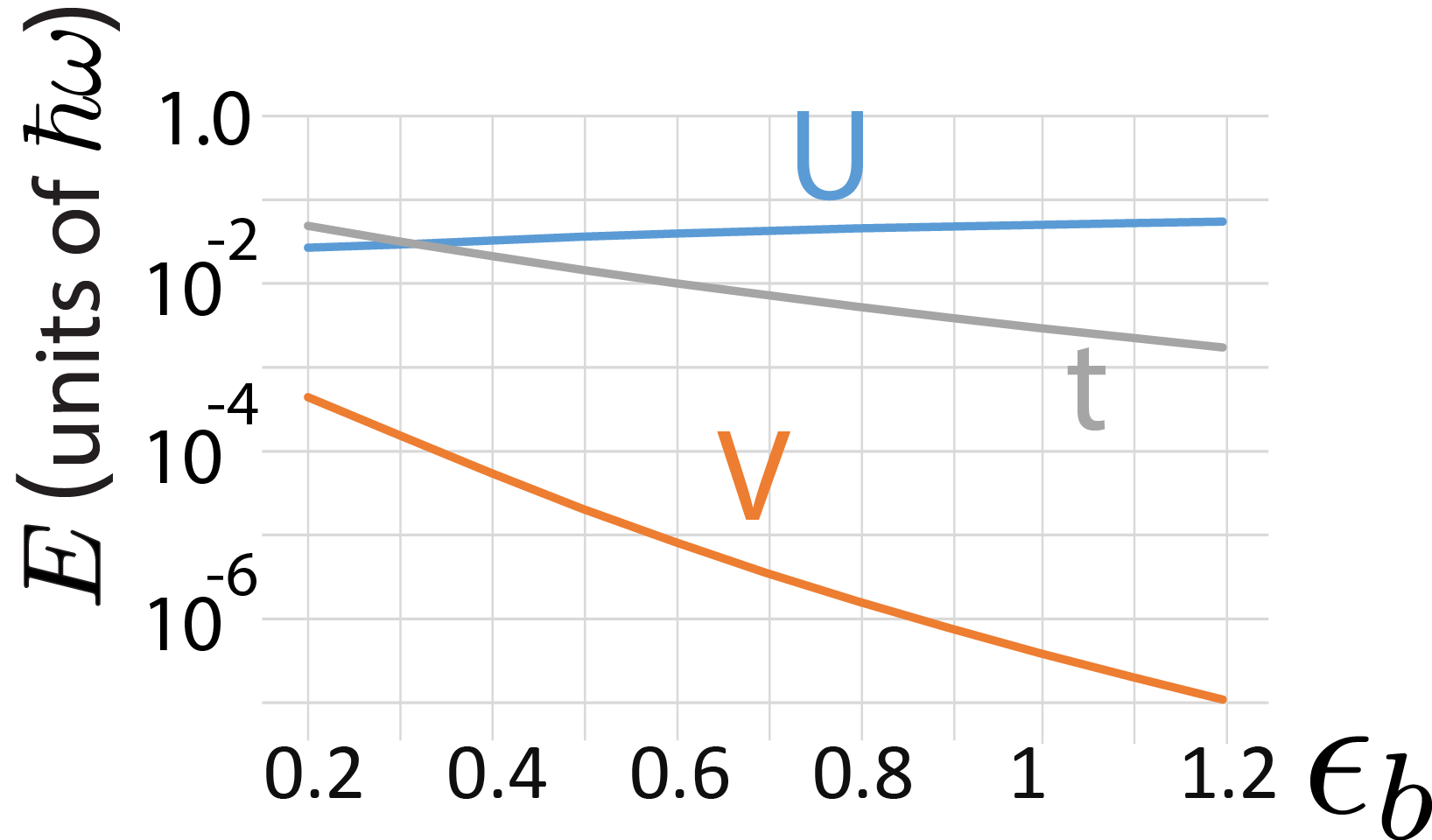}
\caption{
Evolution of the Extended-Hubbard Parameters $U,V,t$ as a function of the inter-well barrier-height ratio 
$\epsilon_b$. The distance between the wells is $d_w=6$ $\mu$m, which results in $V_0=2.68$ kHz. The remaining 
trap parameters are $\hbar\omega = 1$ kHz, $l_0=1.296$ $\mu$m, and $\sigma=0.184$ $\mu$m (note the logarithmic 
energy scale).   
}
\label{fig4_t}
\end{figure}

When $\epsilon_b > 0.4$ \cite{yann15,yann09,yann99}, the maximum value (i.e., $0.045t$) of $\Delta_p$ and the
range $0 \leq U/t < 4.8$, corresponding to $\Delta_p > 0$ (i.e., hole pairing), agree with those found for a 
single-plaquette pure Hubbard model \cite{tsai06,tsai08,schum02}. The additional dependence of $\Delta_p$ on 
$\epsilon_b$ [see Fig.\ \ref{fig3_t}(b)] cannot be described by the standard Hubbard model; it reflects the 
effect of Hamiltonian terms that are present in the CI calculation, but are absent in the standard Hubbard 
model, pointing to possible applications of the extended Hubbard Hamiltonian \cite{fali69,lueh15} in the 
optical traps assemblage. Particularly relevant here is the off-site 
repulsion $V$ which effectively reduces \cite{fali69} the on-site $U$. However, $V$ decreases strongly for 
increasing intersite barrier heights $\epsilon_b$ (see Fig.\ \ref{fig4_t}), whereas 
$U$ is highly insensitive. When $V$ becomes sufficiently small relative to $U$, the standard Hubbard 
single-plaquette results are recovered [see inset in Fig.\ \ref{fig3_t}(b)]. 

\begin{figure}[t]
\centering\includegraphics[width=7.6cm]{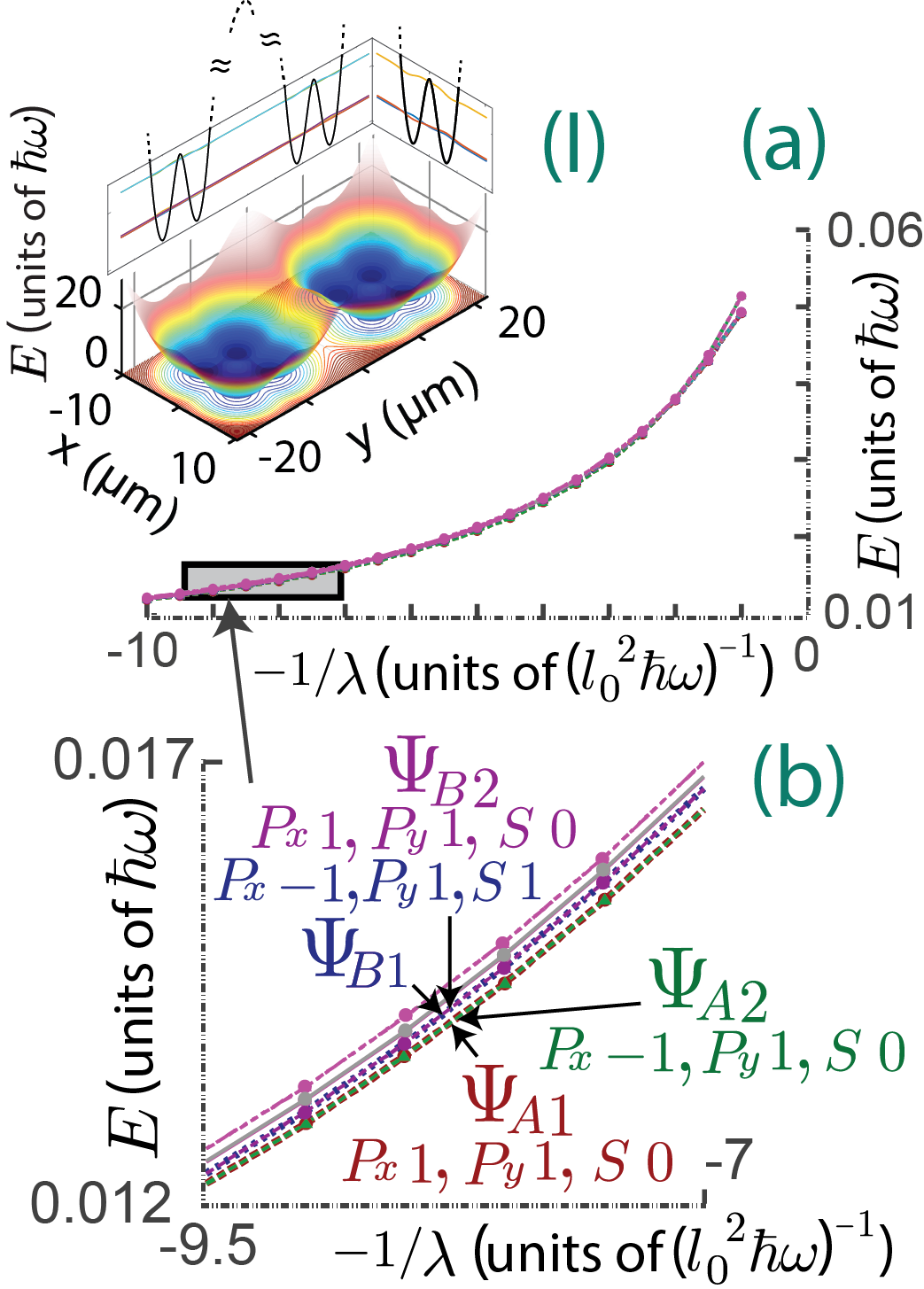}
\caption{
CI spectra for a double plaquette wc-TPM with inter/intra-plaquette distances $D= 18$ $\mu$m and 
$d_w= 6$ $\mu$m, 
(I) Illustration of the external potential for the double plaquette.
(a) Spectrum in the extended interval $-10 < -1/\lambda < -1$.
(b) Enlarged spectrum in the limited interval $-9.5< -1/\lambda < -7$.
}
\label{fig5_t}
\end{figure}

$U$ in Fig.\ \ref{fig4_t} is the on-site 
repulsion as determined from the ground state energy evolution as a function of the interaction strength 
$\lambda$ in a single well. $V$ is the inter-site repulsive interaction, determined through the matrix element of 
$V({\bf r}_i,{\bf r}_j)$ (see Eq.\ \ref{2bi}) taken for the ground state wavefunctions, 
described by Gaussians determined through fitting to the CI SPDs. $t$ is the inter-well tunneling parameter, as 
determined through the energy difference between the two lowest energy states in the $y$ direction (the tunneling 
split). Note the insensitivity of $U$ vs $\epsilon_b$ compared to the strong variation in $V$. 
The tunneling parameter $t$ exhibits a reduced sensitivity compared to $V$. The $V$ term constitutes a perturbation
to the pure Hubbard model. When the $V$ term becomes of the order of magnitude of $\Delta_p$, i.e., $0.05t$, the 
pairing gap vanishes; this happens for values of $\epsilon_b \lesssim 0.3$ for the plaquette considered in this 
figure. 

The role played by the off-site  interaction $V$ is further illustrated in Figs.\ \ref{fig3_t}(c,d), where 
$\Delta_p$ is plotted for the case when the intersite distance in the single plaquette is
$d_w=2.5$ nm [compared with $d_w=6$ nm in Fig.\ \ref{fig3_t}(a,b)]. Clearly much higher values of the intersite 
potential barriers (i.e., $\epsilon_b > 6$) are required to reach the $\Delta_p > 0$ region.

\section{Weakly-coupled two-plaquette molecule}
\label{dp}

The pairing-gap behavior estimated from a single 4-well optical plaquette should be reflected in the 
properties of a two-plaquette molecule (TPM) when the two plaquettes are weakly coupled (wc). For the single 
plaquette [Eq.\ (\ref{dgap}) and Fig.\ \ref{fig3_t}, $\Delta_p$ indicates that the GS of the wc-TPM
is associated with a (left,right) (4,2) or a (2,4) particle 
distribution, with the equal-particle arrangement between the (left, right) plaquettes, i.e., (3,3), 
corresponding to an excited state; $\Delta_p < 0$ indicates reverse energy ordering. The GS and lowest excited
state wave functions of the wc-TPM show complex behavior corresponding to entangled two mirror-reflected
charge-density-wave (CDW) or spin-density-wave (SDW) symmetry-broken configurations. 
Experimental probing and quantitative analysis of such entangled states has recently been demonstrated 
\cite{joch15.2}, based on inducing particle escape from the optical wells by lowering one 
side of the trapping potential.   

\begin{figure*}[t]
\centering\includegraphics[width=16cm]{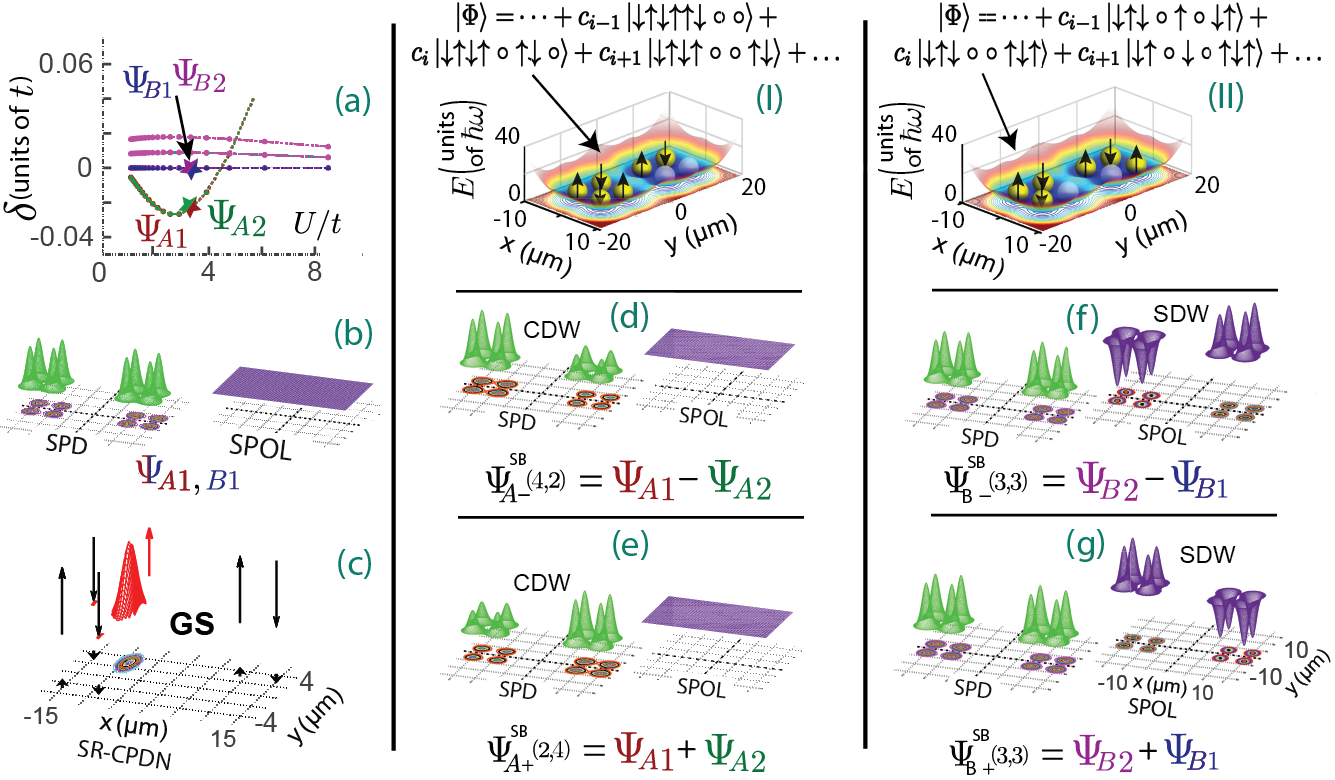}
\caption{
CI wave functions for a double plaquette wc-TPM with inter/intra-plaquette distances 
$D= 18$ $\mu$m and $d_w= 6$ $\mu$m. 
(a) Subtracted spectrum vs. $U/t$. $\delta<0$ indicates hole pair-binding. 
(b) SPD and SPOL for the two lowest states ${\it A}1$ and ${\it A}2$ at the point marked
by the stars in (a). 
(c) SR-CPDN for the GS ${\it A}1$ in (a). The observation points (black arrows) form (together 
with the predicted probability peak (spin-up, red) a (4,2) configuration. 
(d,e) The SPD and SPOL of the symmtery broken states in the ${\it A}$ manifold exhibiting formation of a CDW.
(f,g) Same as (d,e) for the SB states of the ${\it B}$ manifold exhibiting formation of a SDW. 
}
\label{fig6_t}
\end{figure*}

Fig.\ \ref{fig5_t} displays CI results for the spectra of $N=6$ $^6$Li atoms
in two weakly-coupled plaquettes separated by $D=18$ $\mu$m. The well-separation within the identical 
plaquettes is $d_w=6$ $\mu$m [Fig.\ \ref{fig5_t}(I)], so that 
the intraplaquette barriers are sufficiently high to yield a vanishing off-site $V$, see above.
The corresponding low-energy spectrum is plotted versus $-1/\lambda$; the entire spectrum is plotted in 
Fig.\ \ref{fig5_t}(a) and an enlarged view of the  spectrum in the interval  $-9.5 < 1/\lambda < -7$ is 
displayed in Fig.\ \ref{fig5_t}(b). We focus on the two (lowest) pairs of degenerate states denoted as 
(${\it A}1$, ${\it A}2$) and (${\it B}1$, ${\it B}2$) in Fig.\ \ref{fig5_t}(b). 

The CI method preserves all the quantum numbers, which are explicitly denoted in Fig.\ \ref{fig5_t};
i.e., ($\pm 1$) associated with the parities $P_x$ and $P_y$ along the $x$ and $y$ directions, and the 
total spin values $S=0$ (for ${\it A}1$, ${\it A}2$, ${\it B}2$) or $S=1$ (for ${\it B}1$).

The pairs ${\it A}$ and ${\it B}$ constitute the four lowest-in-energy states in the whole range 
$-10 < -1/\lambda < -1$ that we investigated [Fig.\ \ref{fig5_t}(b)]. The crossing at 
$-1/\lambda_c = -2.8$ is reflected in the modified spectrum generated by subtracting the energy 
$E({\it B}1)$ at each point $-1/\lambda$; see the pocket in Fig.\ \ref{fig6_t}(a), 
where the horizontal axis is expressed now in units of
$U/t$ and the crossing occurs at $U/t=4.8$. For $0 < U/t < 4.8$, the ${\it A}$ pair is lower in energy, 
with the $P_x=1$, $P_y=1$, and $S=0$ state (denoted as ${\it A}1$) being the ground state; for $4.8< U/t$,
the ${\it B}$ pair is lower in energy, with the $P_x=1$, $P_y=1$, and $S=1$ state (denoted as ${\it B}1$) 
being the ground state. 

To investigate the microscopic structure of the states in the ${\it A}$ and ${\it B}$ pairs, we first 
display the CI SPDs [$\rho({\bf r},\uparrow)+\rho({\bf r},\downarrow)$] and spin-polarization 
densities [SPOLs, $\rho({\bf r},\uparrow)-\rho({\bf r},\downarrow)$] for the states ${\it A}1$ and 
${\it B}1$ in Fig.\ \ref{fig6_t}(b); see definitions in Appendix \ref{defs}.
These CI SPDs and SPOLs are identical for both states, since both 
parities $P_x$ and $P_y$, as well as the total spin $S$, are preserved. In particular, the SPDs exhibit eight
humps of equal height, integrating to a total number of $N=6$ particles. Furthermore, the SPOLs are 
structureless plane surfaces. The SPDs and SPOLs for the CI states ${\it A}2$ and ${\it B}2$ exhibit a 
similar behavior.

\begin{figure}[t]
\centering\includegraphics[width=8cm]{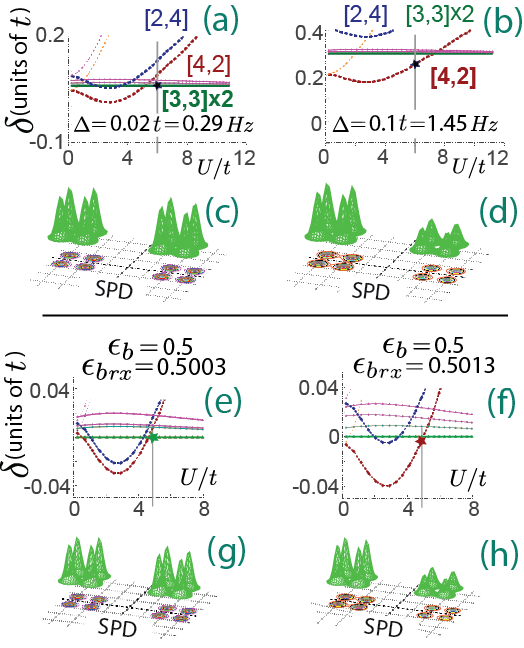}
\caption{
(a-d) Symmetry breaking by applying a tilt $\Delta$, or (e-h) by distorting the internal structure of one of 
the plaquettes. For the chosen value of $U/t$ in (a) with $\Delta=0.02t$, no-pair-binding is observed as evident 
from the SPD in (c). However, for a larger tilt, $\Delta = 0.1 t$, the spectrum in (b) and SPD in (d) indicate GS 
pair-binding, reflecting a (4,2) configuration in (k).  
Similarly, changing the intra-plaquette energy barrier in the right plaquette from $\epsilon_b = 0.5$ to 
$\epsilon_{brx} = 0.5003$ (e) results in no pair binding [see also SPD in (g)] for the chosen value of $U/t$, 
whereas a slightly larger barrier, $\epsilon_{brx} = 0.5013$ (f) induces pair-binding as 
evident from the spectrum in (f) and SPD in (h).
}
\label{fig7_t}
\end{figure}

Direct insight into the nature of the GS structure is afforded by the SR-CPDN shown in Fig.\ \ref{fig6_t}(c), 
where the appearance of a peak in the left plaquette (see red peak), integrating to very-close to unit 
probability for finding an up-spin particle there when fixing the positions and spins of the other five atoms 
(see black arrows), indicates a dominant (4,2) wc-TPM configuration. Little double occupancy is portrayed by small
red peaks on diagonal sites (left plaquette).    

Although the CI-calculated densities appear uniform across the two plaquettes [see Fig.\ \ref{fig6_t}(b), the 
microscopic structures of the ${\it A}$ and ${\it B}$ states are different: 
they involve different superpositions of many primitive basis functions, 
$\Omega_i$, as exemplified by the $|\Phi\rangle$ superpositions in Figs.\ 
\ref{fig6_t}(I) and \ref{fig6_t}(II). These superpositions contain 
also primitives with doubly-occupied sites (denoted as $d$ earlier, not shown explicitly). For illustrative 
purposes, the pictorial representations in these panels describe visually one of these primitives associated 
with the (4,2) [Fig.\ \ref{fig6_t}(I)] and the (3,3) [Fig.\ \ref{fig6_t}(II)] configurations.
 
As aforementioned, detailed information can be extracted from the CI wave functions with the help of
the SR-CPDs \cite{yann15,yann16}. However, for $N=6$ fermions in $M=8$ sites, calculating the large number of
needed CPDs is time consuming. Nevertheless, key features of the hidden anatomy of the entangled states in 
the ${\it A}$ and ${\it B}$ pairs can be revealed through the implementation of a forced breaking of their 
$P_x$ parity symmetry, i.e., by subtracting and adding their member states and constructing the four 
auxiliary states ${\it A}^{SB}_\mp = {\it A}1 \mp {\it A}2$ and 
${\it B}^{SB}_\mp = {\it B}1 \mp {\it B}2$. 

The expectation value of $P_x$ in the ${\it A}^{SB}_\mp$ and ${\it B}^{SB}_\mp$ states vanishes; this 
symmetry-breaking results in different characteristics of the above two symmetry-broken states.  
The SPD for ${\it A}^{SB}_-$ exhibits a (4,2) particle configuration [Fig.\ \ref{fig6_t}(d), left],
while the associated SPOL is featureless [Fig.\ \ref{fig6_t}(d), right]. The SPD for ${\it A}^{SB}_+$ 
corresponds to the mirror-reflected (2,4) configuration [Fig.\ \ref{fig6_t}(e), left], with the associated 
SPOL remaining structureless [Fig.\ \ref{fig6_t}(e), right].
The two fermions (or two holes) in the two-particle plaquette are not localized on specific sites. Indeed 
after the symmetry breaking, the wave functions are still an entangled superposition of many $\Omega_i$'s, 
thus distributing the particles (or holes) with equal probability over all sites in each 
plaquette. In a plaquette with two fermions, the volumes under the four density humps integrate to
$N = 2$, whereas the volumes under the four density humps in a 4-fermion plaquette integrate to $N = 4$.
Naturally, the two broken-symmetry states ${\it A}^{SB}_\mp$ can be 
characterized as a charge density wave (CDW), because they exhibit a modulation in the SPDs, but none in the 
spin polarizations.

The SPDs and spin polarizations of the ${\it B}^{SB}_\mp$ states are displayed in Figs.\ \ref{fig6_t}(f) and
\ref{fig6_t}(g). In both cases, the SPDs are symmetric with respect to the left and right plaquettes. 
Furthermore, the volumes under the humps in each plaquette integrate to $N=3$, indicating that these
states have a (3,3) configuration. However, the left-right asymmetry (due to the broken $P_x$ symmetry) 
emerges now as an asymmetry in the spin polarization. The three fermions (with total spin $S=1/2$) in one
plaquette have a spin peojection $S_z=\pm 1/2$, while the remaining three fermions in the other plaquette 
have the opposite spin projection $S_z=\mp 1/2$. The ${\it B}^{SB}_\mp$ broken-symmetry states 
exhibit spin-density wave (SDW) characteristics, see right panels in Figs.\ \ref{fig6_t}(f,g).

\begin{figure}[t]
\centering\includegraphics[width=8.0cm]{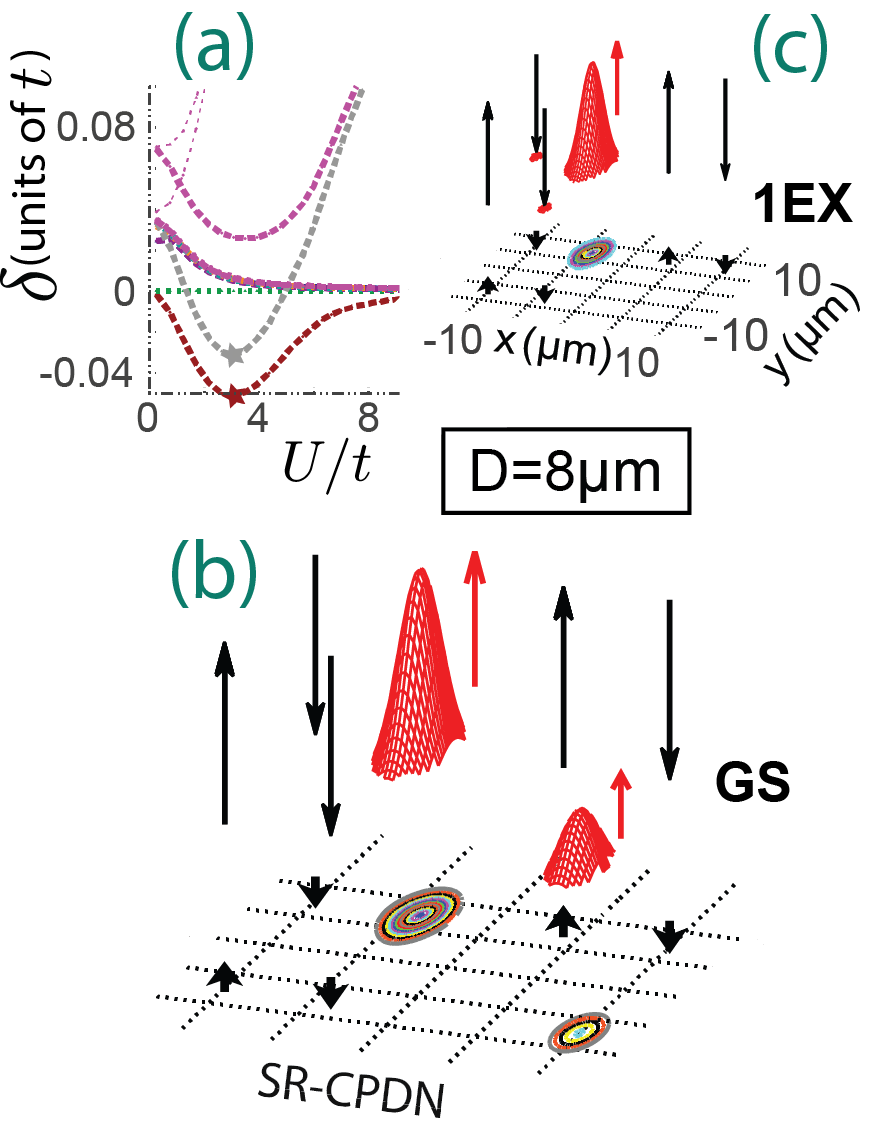}
\caption{
(a) CI spectrum for a double plaquette with inter/intra-plaquette distances $D= 8$ $\mu$m and $d_w = 6$ $\mu$m 
showing degeneracy splitting between the GS and 1EX [compare with Fig.\ \ref{fig6_t}(b), $D=18$ $\mu$m]. 
(b,c) The SR-CPDNs corresponding to the GS and 1EX in (a) (see stars respectively) were evaluated for $U/t = 3.5$. 
The SR-CPDN observation points are marked by black arrows with indicated spin directions. The predicted SR-CPDNs 
are denoted in red. 
}
\label{fig8_t}
\end{figure}

In addition to the above analysis, the (4,2) or (2,4) components of the ${\it A}$ states, but not
the (3$\uparrow$,3$\downarrow$) or (3$\downarrow$,3$\uparrow$) components of the ${\it B}$ states can be 
separated in actual experiments by lifting the left/right degeneracy with the help of two processes: (1) 
Tilting one plaquette with respect to the other [Fig.\ \ref{fig7_t}(a-d)], and 
(2) distorting one of the two plaquettes  Fig.\ \ref{fig7_t}(e-h)].
From the CI spectrum depicted in Fig.\ 
\ref{fig7_t}(b), the degeneracy of the states in the {\it A} pair is lifted for a tilt $\Delta = 0.1 t$, with 
one of the states becoming the GS at $U/t = 6$ (see vertical bar). Moreover the 
corresponding CI SPD plotted in Fig.\ \ref{fig7_t}(d) shows that the tilt induced the symmetry breaking discussed
earlier, i.e., a (4.2) configuration; for smaller tilt results, see Figs.\ \ref{fig7_t}(a,c). Similar 
symmetry-breaking results (i.e., spectral energy gap corresponding to the emergence of a (3,3), CDW 
configuration) are found also for distortions of the plaquette-landscape [see Figs.\ \ref{fig7_t}(e-h)].
 
While we focused here on conditions allowing exploration of pure interplaquette entanglement effects, the 
sensitivity of the energy spectrum and entanglement characteristics to the interplaquette distance (i.e.,
increasing tunneling between the plaquettes) is illustrated in Fig.\ \ref{fig8_t}(a-c), where the spectrum for a 
double-plaqutee with $d_w = 6$ $\mu$m and $D = 8$ $\mu$m is displayed.
The spectrum in Fig.\ \ref{fig8_t}(a) shows formation of an energy gap between the two lowest states
[compare with Fig.\ \ref{fig6_t}(a)]. The SR-CPDN in Fig.\ \ref{fig8_t}(b), corresponding to the GS at 
$U/t = 3.5$ (see red star), reflects  contributions from both the (4,2) and (3,3) configurations 
(see the large red peak in the left plaquette and a smaller one in the right plaquette, corresponding to the 
probabilities of finding spin-up atoms in these locations). On the other hand the result [Fig.\ \ref{fig8_t}(c)]
for the 1EX state [gray star in Fig.\ \ref{fig8_t}(a)] portrays formation of a pure (4,2)$-$(2,4) configuration 
[compare to the (4,2)$+$(2,4) GS of the wc-TPM in Fig.\ \ref{fig6_t}(c); the small difference between the $+$
and $-$ cases is not visible in the plots].

\section{Summary}
\label{summ}

In this work we have developed and implemented a configuration-interaction-based computational methodology for 
obtaining exact solutions to the microscopic many-body quantum Hamiltonian describing ulracold fermionic atoms 
(here $^6$Li atoms) moving under the influence of an optically-induced multi-well confining potential surface and 
with short-range interatomic repulsive interactions; we remark here that the same CI methodology can be extended 
in a straightforward manner to treat interatomic dipolar interactions, or confined electrons interacting via 
long-range Coulomb interactions. A similar type of potential energy surface, made of the assembly of plaquettes  
-- each comprising four sites (potential wells) arranged in a square geometry --  has been introduced in early 
investigations \cite{scal96,tsai06} of electronic high-temperature superconductivity in cuprate materials, in the 
context of the two-dimensional Hubbard model often mentioned as a starting point for formulating a theoretical 
understanding of unconventional superconductivity \cite{lee06,lee08}. 

The Hubbard model has been originally derived for the description of the behavior of strongly correlated electrons in 
solids \cite{hubb78}. However, variants of this model have 
also been implemented, for over a decade, targeting investigations of 
strongly-interacting ultracold atoms (bosonic or fermionic) in optical lattices, with the interest in such systems 
originating from remarkable advances in atom cooling and optical trapping techniques, that opened heretofore untapped 
prospects of preparing, emulating, and measuring the behavior of strongly-interacting quantum many-body systems under 
pristine, defect-free, environments. Analysis of such ultracold-atom emulations of interacting quantum many-body 
systems revealed on various occasions \cite{lueh15} that many Hubbard models that were simulated with ultracold atoms 
were found not to have a standard form -- meaning that the corresponding Hamiltonians, required for analysis of these 
results, frequently had to incorporate terms that are not included in the standard Hubbard model and its customary 
variants. These extra terms may include correlated and occupation-dependent tunneling within the lowest band, as well 
as correlated tunneling and occupation of higher bands \cite{lueh15}. 

The above findings connote that progress in developing future ultracold-atom emulations of interacting quantum 
many-body systems (including quantum magnetism and the origins of unconventional pairing mechanisms), and theoretical 
methods for the analysis of such emulations (including the development and implementation of effective models, such as
extended and non-standard Hubbard Hamiltonians), require benchmark exact calculations (such as the ones introduced in
this work), which owing to their ab-initio nature incorporate (with no restriction or approximation) all contributions
to the many-body microscopic Hamiltonian. With such computational methodology, one can then proceed to construct, 
assess, and improve effective models for a reliable analysis of the new forthcoming data. Indeed, in this work we 
focused on providing benchmark exact quantum-mechanical solutions, aiding and enabling a bottom-up approach aimed at 
ultracold atoms experiments, and their analysis, starting from a single ultracold atom plaquette as a building block, 
and  progressing in a systematic manner to double plaquettes (with variable inter-plaquette couplings) and larger, 
more complex, systems.

Our exact Schr\"{o}dinger-equation ultracold-fermionic-atoms-plaquette simulations demonstrated entangled 
$d$- and $s$- wave RVB states coexisting with partial double occupancies, uncovered hole-pairing phase diagrams, 
and explored the robustness of the energy spectrum and entanglement of a double-plaquette system (assembled from 
entangled-multipartite-plaquettes) through symmetry-breaking via interplaquette tilting and distortion and as a 
function of the strength of interplaquette coupling. These calculations may serve as a launchpad for further exact 
calculations, the development of approximate treatments \cite{baer09,yoko06,mezz16}, and experiments 
(including the use of site resolved microscopy enabling direct observation of charge and spin correlations
\cite{bloc16,grei16,zwie16}) on systems with hierarchically increasing complexity \cite{rega16,barr16,endr16}.

\newpage
\begin{acknowledgments}
Acknowledgment:
Work supported by the Air Force Office of Scientic Research under Award No. FA9550-15-1-0519. 
Calculations were carried out at the GATECH Center for Computational Materials Science.
We thank R. Schumann for correspondence concerning the analytic Hubbard-model solutions of the 4-site plaquette.
\end{acknowledgments}

\appendix

\section{The configuration-interaction method}
\label{cime}

As aforementioned, we use the method of configuration interaction for determining the solution of the many-body 
problem specified by the $N$-fermion general Hamiltonian
\begin{equation}
{\cal H}_{\rm MB}=\sum_{i=1}^{N} H(i) +
\sum_{i=1}^{N} \sum_{j>i}^{N} V({\bf r}_i,{\bf r}_j),
\label{mbhd}
\end{equation}
where ${\bf r}_i$, ${\bf r}_j$ denote the vector positions of the $i$ and $j$ fermions (e.g., $^6$Li atoms). 
This Hamiltonian is the sum of a single-particle part $H(i)$, which includes a kinetic  energy term and a 
single-particle external confinement potential (see Fig.\ \ref{fig1_t}) that expresses formation of an 
interwell barrier between the individual wells, and the two-particle interaction $V({\bf r}_i,{\bf r}_j)$.

For the case of 2D ultracold atoms, the two-body repulsion is taken as a Gaussian given by Eq.\ (\ref{2bi}).

In the CI method, one writes the many-body wave function 
$\Phi^{\text{CI}}_N ({\bf r}_1, {\bf r}_2, \ldots , {\bf r}_N)$ as a linear
superposition of Slater determinants 
$\Psi^N({\bf r}_1, {\bf r}_2, \ldots , {\bf r}_N)$ that span the many-body
Hilbert space and are constructed out of the single-particle 
{\it spin-orbitals\/} 
\begin{equation}
\chi_j (x,y) = \varphi_j (x,y) \alpha, \mbox{~~~if~~~} 1 \leq j \leq K,
\label{chi1}
\end{equation}
and 
\begin{equation}
\chi_j (x,y) = \varphi_{j-K} (x,y) \beta, \mbox{~~~if~~~} K < j \leq 2K, 
\label{chi2}
\end{equation}
where $\alpha (\beta)$ denote up (down) spins. Namely
\begin{equation}
\Phi^{\text{CI}}_{N,q} ({\bf r}_1, \ldots , {\bf r}_N) = 
\sum_I C_I^q \Psi^N_I({\bf r}_1, \ldots , {\bf r}_N),
\label{mbwf}
\end{equation}
where 
\begin{equation}
\Psi^N_I = \frac{1}{\sqrt{N!}}
\left\vert
\begin{array}{ccc}
\chi_{j_1}({\bf r}_1) & \dots & \chi_{j_N}({\bf r}_1) \\
\vdots & \ddots & \vdots \\
\chi_{j_1}({\bf r}_N) & \dots & \chi_{j_N}({\bf r}_N) \\
\end{array}
\right\vert,
\label{detexd}
\end{equation}
and the master index $I$ counts the number of arrangements 
$\{j_1,j_2,\ldots,j_N\}$ under the restriction that 
$1 \leq j_1 < j_2 <\ldots < j_N \leq 2K$. Of course, $q=1,2,\ldots$ counts
the excitation spectrum, with $q=1$ corresponding to the ground state.

The many-body Schr\"{o}dinger equation 
\begin{equation}
{\cal H}_{\rm MB} \Phi^{\text{EXD}}_{N,q} = E^{\text{EXD}}_{N,q} \Phi^{\text{EXD}}_{N,q}
\label{mbsch}
\end{equation}
transforms into a matrix diagonalizatiom problem, which yields the coefficients 
$C_I^q$ and the eigenenergies $E^{\text{CI}}_{N,q}$. Because the
resulting matrix is sparse, we implement its numerical diagonalization 
employing the well known ARPACK solver \cite{arpack}.

The matrix elements $\langle \Psi_N^{I} | {\cal H} | \Psi_N^{J} \rangle$
between the basis determinants [see Eq.\ (\ref{detexd})] are calculated using
the Slater rules \cite{szabobook}. Naturally, an important ingredient in this
respect are the matrix elements of the two-body interaction,
\begin{equation}
\int_{-\infty}^{\infty} \int_{-\infty}^{\infty} d{\bf r}_1 d{\bf r}_2
\varphi^*_i({\bf r}_1) \varphi^*_j({\bf r}_2) V({\bf r}_1,{\bf r}_2)
\varphi_k({\bf r}_1) \varphi_l({\bf r}_2),
\label{clme}
\end{equation}
in the basis formed out of the single-particle spatial orbitals 
$\varphi_i({\bf r})$, $i=1,2,\ldots,K$. 
In our approach, these matrix elements are determined numerically and stored separately.

The Slater determinants $\Psi^N_I$ [see Eq.\ (\ref{detexd})] conserve the
third projection $S_z$,  but not the square $\hat{\bf S}^2$ of the total spin. 
However, because $\hat{\bf S}^2$ commutes with the many-body Hamiltonian, the
CI solutions are automatically eigenstates of $\hat{\bf S}^2$ with eigenvalues
$S(S+1)$. After the diagonalization, these eigenvalues are determined by
applying $\hat{\bf S}^2$ onto $\Phi^{\text{CI}}_{N,q}$ and using the relation
\begin{equation}
\hat{{\bf S}}^2 \Psi^N_I = 
\left [(N_\alpha - N_\beta)^2/4 + N/2 + \sum_{i<j} \varpi_{ij} \right ] 
\Psi^N_I,
\end{equation}
where the operator $\varpi_{ij}$ interchanges the spins of fermions $i$ and 
$j$ provided that their spins are different; $N_\alpha$ and $N_\beta$ denote 
the number of spin-up and spin-down fermions, respectively.

When there is no tilt [i.e., $\Delta = 0$, see Fig.\ \ref{fig1_t}(b)], the $xy$-parity 
operator associated with reflections about the origin of the axes is defined as
\begin{equation}
\hat{\cal P}_{xy} 
\Phi^{\text{CI}}_{N,q}({\bf r}_1, {\bf r}_2, \ldots, {\bf r}_N) =
\Phi^{\text{CI}}_{N,q} (-{\bf r}_1, -{\bf r}_2, \dots, -{\bf r}_N)
\label{xypar}
\end{equation}
and has eigenvalues $\pm 1$. With the separable in $x$ and $y$ basis that we 
use, it is easy to calculate the parity eigenvalues for the Slater 
determinants, Eq.\ (\ref{detexd}), that span the many-body Hilbert space.
The many-body Hamiltonian used in this paper (without an applied magnetic field) conserves
also the partial $\hat{\cal P}_{x}$ and $\hat{\cal P}_{y}$ parities. 

\section{Single particle densities and conditional probability distributions -- SPD, SPOL, SR-CPD2, and SR-CPDN}
\label{defs}

Denoting the CI wave function as $| \Phi^{\text{CI}} \rangle$, the single-particle density (SPD) is defined as
\begin{equation}
n({\bf r}) =
\langle \Phi^{\text{CI}} | \sum_{i=1}^N \delta({\bf r}_i-{\bf r}) | \Phi^{\text{CI}} \rangle.
\label{spdensity}
\end{equation}

Furthermore the single-particle spin polarization (SPOL) is defined as
\begin{equation}
\text{SPOL}({\bf r}) =
\langle \Phi^{\text{CI}} | \sum_{i=1}^N \delta({\bf r}_i-{\bf r})
(\delta_{\uparrow\sigma_i}-\delta_{\downarrow\sigma_i}) | \Phi^{\text{CI}} \rangle.
\label{spoleq}
\end{equation}
where $\uparrow$ and $\downarrow$ denote the up and down values, respectively, of the $\sigma$ spin variable.
The SPOL is the difference between the up and down spin densities. 

We probe the intrinsic structure of the CI eigenstates using the spin resolved conditional probability distribution
$P({\bf r}\sigma,{\bf r}_{0}\sigma_0)$ defined by the expression \cite{yann09,yann07.2}
\begin{eqnarray}
&&P({\bf r}\sigma,{\bf r}_{0}\sigma_0)= ~~~~~~~~~~~~~~~~~~~~~~~~~~~ \nonumber \\
&&\langle \Phi^{\text{CI}} |\sum_{i=1}^{N}\sum_{j \neq i}^{N}
\delta({\bf r}_{i}-{\bf r})\delta({\bf r}_{j}-{\bf r}_{0})\delta_{\sigma\sigma_i}
\delta_{\sigma_o\sigma_j}| \Phi^{\text{CI}} \rangle,
\label{cpd}
\end{eqnarray}
where $\delta$ is the Dirac delta function, $N$ the total number of particles, $i,j$ are particle 
indices, $\sigma_{i}$ and $\sigma_{j}$ represent the spins of particles $i$ and $j$. The position of the fixed 
point is ${\bf r}_0$ and the spin of the fixed particle is $\sigma_0$; $r_0$ (together with the associated spin at 
that location, $\sigma_0$) is referred to as the observation point. $\sigma$ is the spin of the particle whose 
spatial distribution we want to know. Since the spin-resolved conditional probability distribution [in 
Eq.\ (\ref{cpd})] is represented by a two-body operator we denote it as SR-CPD2. 

However, we found that sometimes SR-CPD2s alone are not sufficient to fully decipher the intrinsic configuration 
of the emerging quantum states. In such a case, one needs to calculate higher correlation functions. In this paper,
we use the SR-CPDN ($N$-point correlation function) defined as the modulus square of the full many-body CI wave 
function, i.e., 
\begin{eqnarray}
P({\bf r}\sigma;{\bf r}_1&&\sigma_1,{\bf r}_2\sigma_2,...,{\bf r}_{N-1}\sigma_{N-1})=
~~~~~~~~~~~~~~~~~~~~~~~~ \nonumber\\
&&| \Phi^{\text{CI}} ({\bf r\sigma};{\bf r}_1\sigma_1,{\bf r}_2\sigma_2,...,{\bf r}_{N-1}\sigma_{N-1})|^2,
\label{npoint_def}
\end{eqnarray}
where one fixes the positions and spins of $N-1$ particles and inquiries about the
(conditional) probability of finding the $N$th particle with spin $\sigma$ at any position
${\bf r}$.

\section{Spin eigenfunction determination}
\label{spin}

\subsection{RVB state in a single plaquette}
\label{rvb}

In this section we provide a detailed description of the procedure used to identify the d-wave RVB 
state in our spin resolved SR-CPD2s. 

As first described in Ref.\ \cite{yann09} one can map SR-CPD2s to spin eigenfunctions by analyzing the volumes 
underneath the SR-CPD2s. The versatility of this method has been demonstrated several times 
\cite{yann09,yann16} and detailed explanations of the methodology can be found in 
\cite{lidis09,yann16,yann09}. This procedure was carried out by us to identify the RVB states within our 
plaquettes. The starting point is the general spin eigenfunction for four spin $1/2$ fermions trapped in a 
single plaquette with quantum numbers $S=0,S_z=0$, which is given as \cite{yann09}:
\begin{widetext}
\begin{align}
\begin{split}
\chi_{00}=&\sqrt{\frac{1}{3}}\sin \theta |\uparrow\uparrow\downarrow\downarrow \rangle+
\left(\frac{1}{2}\cos \theta - \sqrt{\frac{1}{12}}\sin \theta\right) 
|\uparrow\downarrow\uparrow\downarrow \rangle\\
&-\left(\frac{1}{2}\cos \theta + 
\sqrt{\frac{1}{12}}\sin \theta\right) |\uparrow\downarrow\downarrow\uparrow \rangle-
\left(\frac{1}{2}\cos \theta + \sqrt{\frac{1}{12}}\sin \theta\right)
|\downarrow\uparrow\uparrow\downarrow \rangle\\
&+\left(\frac{1}{2}\cos \theta - \sqrt{\frac{1}{12}\sin \theta}\right) 
|\downarrow\uparrow\downarrow\uparrow \rangle+
\sqrt{\frac{1}{3}}\sin \theta |\downarrow\downarrow\uparrow\uparrow \rangle
\end{split}
\label{eq:generalspinfun}
\end{align}
\end{widetext}

This function is parameterized by the angle $\theta$ and yields the quantum numbers $S=0,S_z=0$ for all values of 
$\theta$. The angle $\theta$ is a free parameter that for certain values gives spin functions with specified 
characteristics (e.g., the orthogonal functions mentioned below), or can be treated as a ``fitting'' parameter 
(see the discussion below Eq.\ (\ref{eq:partial_conditional_prob}) in connection with $\theta_{CI}$). Eq.\ 
(\ref{eq:generalspinfun}) expresses in a compact form the fact that the dimension of the total-spin space for 
$N=4$ and $S=0$, $S_z=0$ is 2; see the branching diagram in Ref.\ \cite{paunczbook} and Ref.\ 
\cite{yann16}. For example, two orthogonal basis functions can be obtained by setting $\theta=0$ and 
$\theta=\pi/2$. From the general spin function, one can read off the spin components contributing to a 
specific SR-CPD2. For instance, for a spin down SR-CPD2 with a fixed spin-up fermion on position 1 (counting 
from left to right in the corresponding kets), the spin components 
contributing to the conditional probability densities are the following three:

\begin{align}
&\sqrt{\frac{1}{3}}\sin \theta |\uparrow\uparrow\downarrow\downarrow \rangle;\\
&\left(\frac{1}{2}\cos \theta - \sqrt{\frac{1}{12}}\sin \theta\right)
| \uparrow\downarrow\uparrow\downarrow \rangle;\\ 
&\left(\frac{1}{2}\cos \theta + \sqrt{\frac{1}{12}}\sin \theta\right) 
|\uparrow\downarrow\downarrow\uparrow \rangle.
\end{align}

The volume under the hump in position 2 of such a SR-CPD2 is therefore proportional to 

\begin{align}
\begin{split}
&\Pi_{\uparrow\downarrow}(1,2)=\\
&\left(\frac{1}{2}\cos \theta - \sqrt{\frac{1}{12}}\sin 
\theta\right)^2+\left(\frac{1}{2}\cos \theta + \sqrt{\frac{1}{12}}\sin \theta\right)^2
\label{eq:partial_conditional_prob}
\end{split}
\end{align}

This is referred to as the partial conditional probability $\Pi_{\uparrow\downarrow}(1,2)$ (partial because it 
corresponds to a part of the full SR-CPD2, i.e. one peak for a specific spin configuration). Using a normalized 
SR-CPD2. one can directly equate the volume underneath the hump in position 2 to $\Pi_{\uparrow\downarrow}(1,2)$ 
and determine the angle $\theta_{CI}$. Due to the involved squares, this procedure is not necessarily unique. The 
unique solution can be found by comparing another hump (i.e., at position 3) to the corresponding partial 
conditional probability $\Pi_{\uparrow\downarrow}(1,3)$. 

However it is important to note that this procedure is only exact as long as (i) The overlap between sites is 
sufficiently small and (ii) The amount of double occupancy is small.

From Fig.\ \ref{fig2_t}(c,d), it is clear that (i) is fulfilled. To assure that the possible error due to (ii) 
is as small as possible, we minimize 
\begin{align}
\begin{split}
&\delta=(\Pi_{\uparrow\downarrow}(1,2)-\mathit{Vol}(2))^2+\\
&(\Pi_{\uparrow\downarrow}(1,3)-\mathit{Vol}(3))^2+(\Pi_{\uparrow\downarrow}(1,4)-\mathit{Vol}(4))^2,
\end{split}
\end{align}
where $\mathit{Vol}(i)$ represents the volume under the SR-CPD2 hump at position $i$. 

\subsection{$d$-wave RVB determination}

For the ground state SR-CPD2 shown in Fig.\ \ref{fig2_t}, this procedure yielded an angle 
$\theta_{CI}=2.618614\pi$. It is known that the spin function for the $d$-wave RVB is given as 
\cite{nasc12}
\begin{align}
\begin{split}
\chi_{d-RV\!B}=&\frac{1}{2\sqrt{3}}|\uparrow\uparrow\downarrow\downarrow\rangle-
\frac{1}{\sqrt{3}} |\uparrow\downarrow\uparrow\downarrow\rangle+
\frac{1}{2\sqrt{3}}|\uparrow\downarrow\downarrow\uparrow\rangle+\\
&\frac{1}{2\sqrt{3}}|\downarrow\uparrow\uparrow\downarrow\rangle-
\frac{1}{\sqrt{3}}|\downarrow\uparrow\downarrow\uparrow\rangle+
\frac{1}{2\sqrt{3}}|\downarrow\downarrow\uparrow\uparrow\rangle
\end{split}
\end{align}
which corresponds to an angle of $\theta_{d-RV\!B}=\frac{5\pi}{6}$. We can therefore conclude that our CI results 
show the presence of a $d$-wave RVB state with 
\begin{align}
1-\left|\frac{\theta_{CI}-\theta_{d-RV\!B}}{\theta_{d-RV\!B}}\right|=99.98\%
\end{align}
fidelity. However while the CI result shows the presence of an RVB with high fidelity, the SR-CPD2 also 
undoubtedly shows the presence of double occupancy. The volume of the hump at position 1 amounts to $10.39\%$. 
This might seem at first contradictory but is in fact the correct solution for an interaction strength of 
$U/t=2.461$. To show this we calculated the ground state solution of the Hubbard Model
\begin{align}
H=-t\sum_{<i,j>,\sigma}(c_{i,\sigma}^\dagger c_{j,\sigma}+c_{j,\sigma}^\dagger c_{i,\sigma})+U\sum_{i=1}^N n_{i\uparrow}n_{i\downarrow}
\end{align}
\noindent which is given as 
\begin{widetext}
\begin{align}
\begin{split}
\psi_{gs}&=
0.455\big(|\uparrow,\downarrow,\uparrow,\downarrow\rangle+|\uparrow,\downarrow,\uparrow,\downarrow\rangle\big)\\
&+0.228\big(-|\uparrow,\uparrow,\downarrow,\downarrow\rangle-|\uparrow,\downarrow,\downarrow,\uparrow\rangle-
|\downarrow,\uparrow,\uparrow,\downarrow\rangle-|\downarrow,\downarrow,\uparrow,\uparrow\rangle\big)\\
&+0.149\big(|\circ,\downarrow,\uparrow,\uparrow\downarrow\rangle
-|\circ,\uparrow,\downarrow,\uparrow\downarrow\rangle\big)+
0.149\big(|\uparrow,\downarrow,\circ,\uparrow\downarrow\rangle-
|\downarrow,\uparrow,\circ,\downarrow\uparrow\rangle\big)\\
&+0.149\big(|\downarrow,\circ,\uparrow\downarrow,\uparrow\rangle-
|\uparrow,\circ,\uparrow\downarrow,\downarrow\rangle\big)+
0.149\big(|\uparrow,\downarrow,\uparrow\downarrow,\circ\rangle-
|\downarrow,\uparrow,\uparrow\downarrow,\circ\rangle\big)\\
&+0.149\big(|\downarrow,\uparrow\downarrow,\circ,\uparrow\rangle-
|\uparrow,\uparrow\downarrow,\circ,\downarrow\rangle\big)+
0.149\big(|\circ,\uparrow\downarrow,\uparrow,\downarrow\rangle-
|\circ,\uparrow\downarrow,\downarrow,\uparrow\rangle\big)\\
&+0.149\big(|\uparrow\downarrow,\circ,\downarrow,\uparrow\rangle-
|\uparrow\downarrow,\circ,\uparrow,\downarrow\rangle\big)+
0.149\big(|\uparrow\downarrow,\uparrow,\downarrow,\circ\rangle-
|\uparrow\downarrow,\downarrow,\uparrow,\circ\rangle\big)\\
&+0.078\big(-|\circ,\circ,\uparrow\downarrow,\uparrow\downarrow\rangle+
|\circ,\uparrow\downarrow,\uparrow\downarrow,\circ\rangle+
|\uparrow\downarrow,\circ,\circ,\uparrow\downarrow\rangle-
|\uparrow\downarrow,\uparrow\downarrow,\circ,\circ\rangle\big)\\
&\;\;\;\;\;=0.79\chi_{d-RV\!B}+0.38\chi_{do}
\end{split}
\label{eq:hubbard_dwave_rvb}
\end{align}
\end{widetext}

The first two lines of Eq.\ (\ref{eq:hubbard_dwave_rvb}) are the $d$-wave RVB component and appear with 
dominant coefficients, however the wavefunction clearly contains contributions from doubly occupied states. 
Summing the squared coefficients of the  spin primitives that contain doubly occupied sites yields a double 
occupancy of $10.17\%$ in excellent agreement with the CI result.

\subsection{s-wave RVB determination}

The same analysis can be performed for the $s$-wave RVB state. The spin function for an $s$-wave RVB state is 
given in literature as\cite{nasc12,rey09}:
\begin{align}
\chi_{s-RV\!B}=\frac{1}{2}\left(|\uparrow\uparrow\downarrow\downarrow\rangle+
|\downarrow\downarrow\uparrow\uparrow\rangle-|\downarrow\uparrow\uparrow\downarrow\rangle-
|\uparrow\downarrow\downarrow\uparrow\rangle\right)
\end{align}

This corresponds to an angle of $\theta_{s-RV\!B}=-\frac{2\pi}{3}$. The angle determined from our CI is 
$\theta_{CI}=-0.66632\pi$ corresponding to a fidelity of 
\begin{align}
1-\left|\frac{\theta_{CI}-\theta_{s-RV\!B}}{\theta_{s-RV\!B}}\right|=99.95\%
\end{align}

However just like the ground state SR-CPD2, the SR-CPD2 for the second excited state shows a non-zero double
occupancy. Therefore we computed the Hubbard model solution for the 2nd excited state at $U/t=2.461$:
\begin{widetext}
\begin{align}
\begin{split}
\psi_{2nd\; exc.}&=
0.361\big(-|\uparrow,\uparrow,\downarrow,\downarrow\rangle+|\uparrow,\downarrow,\downarrow,\uparrow\rangle+
|\downarrow,\uparrow,\uparrow,\downarrow\rangle-|\downarrow,\downarrow,\uparrow,\uparrow\rangle\big)\\
&+0.185\big(-|\uparrow\downarrow,\circ,\uparrow\downarrow,\circ\rangle-
|\circ,\uparrow\downarrow,\circ,\uparrow\downarrow\rangle\big)\\
&+0.153\big(|\circ,\downarrow,\uparrow,\uparrow\downarrow\rangle
-|\circ,\uparrow,\downarrow,\uparrow\downarrow\rangle\big)+
0.153\big(-|\uparrow,\downarrow,\circ,\uparrow\downarrow\rangle+
|\downarrow,\uparrow,\circ,\downarrow\uparrow\rangle\big)\\
&+0.153\big(|\downarrow,\circ,\uparrow\downarrow,\uparrow\rangle-
|\uparrow,\circ,\uparrow\downarrow,\downarrow\rangle\big)+
0.153\big(-|\uparrow,\downarrow,\uparrow\downarrow,\circ\rangle+
|\downarrow,\uparrow,\uparrow\downarrow,\circ\rangle\big)\\
&+0.153\big(|\downarrow,\uparrow\downarrow,\circ,\uparrow\rangle-
|\uparrow,\uparrow\downarrow,\circ,\downarrow\rangle\big)+
0.153\big(-|\circ,\uparrow\downarrow,\uparrow,\downarrow\rangle+
|\circ,\uparrow\downarrow,\downarrow,\uparrow\rangle\big)\\
&+0.153\big(|\uparrow\downarrow,\circ,\downarrow,\uparrow\rangle-
|\uparrow\downarrow,\circ,\uparrow,\downarrow\rangle\big)+
0.153\big(-|\uparrow\downarrow,\uparrow,\downarrow,\circ\rangle+
|\uparrow\downarrow,\downarrow,\uparrow,\circ\rangle\big)\\
&+0.093\big(-|\circ,\circ,\uparrow\downarrow,\uparrow\downarrow\rangle-
|\circ,\uparrow\downarrow,\uparrow\downarrow,\circ\rangle-
|\uparrow\downarrow,\circ,\circ,\uparrow\downarrow\rangle-
|\uparrow\downarrow,\uparrow\downarrow,\circ,\circ\rangle\big)\\
&\;\;\;\;\;=0.72\chi_{s-RV\!B}+0.48\chi_{do}
\end{split}
\label{eq:hubbard_swave_rvb}
\end{align}
\end{widetext}

This eigenfunction predicts a double occupancy on site 1 of $14.52\%$. The volume of the hump on position 1 indicating double occupancy in our CI amounts to $15.78\%$ in good agreement with the Hubbard result.


\newpage

\begin{center}
{\bf SUPPLEMENTAL MATERIAL}\\
(References are renumbered; see list at the end)
\end{center}

\section{External Confining Potentials for Plaquettes}

\subsection{The external potential for a single plaquette}
\label{sec:single_plaquette_hamiltonian}

For our single plaquette calculations we combine the double well external potentials from 
\cite{li_artificial_2009,brandt_double-well_2015,yannouleas_ultracold_2016} in both the $x$ and $y$ directions. 

\begin{align}
v(x_i,y_i)=v_{dw}(x_i)+v_{dw}(y_i),
\end{align}

\noindent where $v_{dw}$ is the double well external potential (see Fig.\ 1 in the main paper). This results in 
a two-dimensional confining potential that has four minima and smooth connecting necks between them. We have 
full control over the distances between the minima in $x$ and $y$ and over the barriers between them. The set of
parameters used to control the double-well potential are $V_b$, $\Delta$, and $d_w$, where $V_b$ is the height of
the smooth inter-well barrier, $\Delta$ is the tilt between the left and right well, and $d_w$ is the distance 
between the wells \cite{li_artificial_2009,brandt_double-well_2015,yannouleas_ultracold_2016}. 
For a graphical illustration of the double-well potential, including the definitions of $V_b$, 
$\epsilon_b = V_b/V_0$, and $V_0$, see Fig.\ 1 of the main paper.

\subsection{The external potential for the double (multi) plaquette}

\begin{figure}[t]
\includegraphics[width=6.5cm]{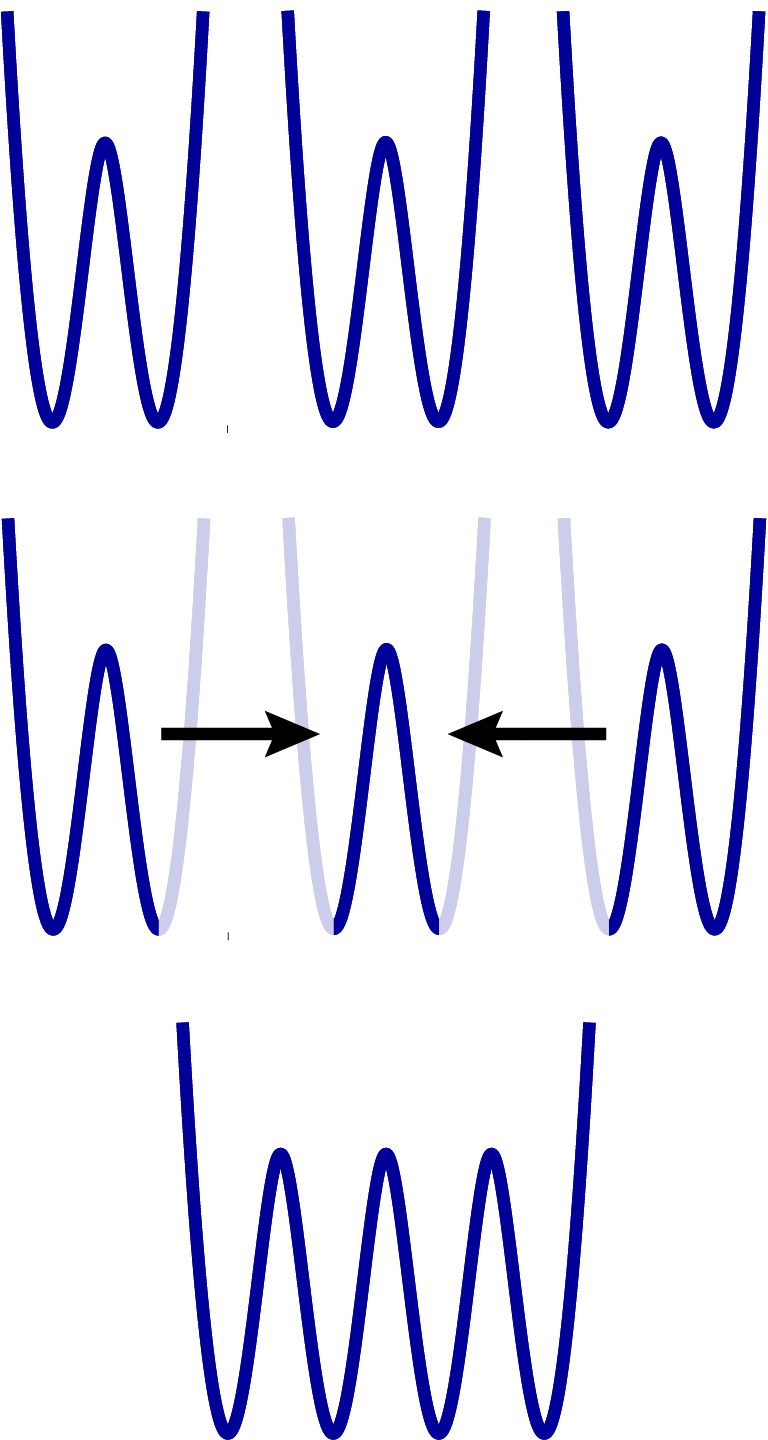} 
\caption{{\bf Schematic of the assemblage of a four-well potential.}
Schematic of the assemblage of a four-well potential. Starting from three separate double-well potentials 
(top row), the inward facing potential walls are removed and the resulting pieces are connected (middle row). 
This creates a four-well potential with smooth necks where we have full control over the individual barriers, 
the distance of the minima and the tilts between the wells. 
} 
\label{fig:shematicquadruplewellassembly}
\end{figure}

For our double-plaquette calculations we use a repetition of the single plaquette external potential from Sec.\
\ref{sec:single_plaquette_hamiltonian}. We keep the potential along $y$ as is. For the potential along $x$ we 
start out with three double-well potentials and use Heaviside theta functions $\Theta$ to cut the inward facing 
potential walls to the right and/or left of the minima. We then combine the resulting potentials. This procedure 
is illustrated in Fig.\ \ref{fig:shematicquadruplewellassembly}. The above procedure 
applies directly to large inter-plaquette distances. 
For smaller inter-plaquette distances end-effects of the two-plaquette assembly are alleviated through 
"symmetrization" of the trapping potential at the left and right edges of the double-plaquette assembly so that it
takes the shape of the intraplaquette interwell potential from the bottom of the well to the midpoint between 
sites.

Just like in Sec.\ \ref{sec:single_plaquette_hamiltonian} this gives us full control over the positions of the 
minima in $x$ and $y$ and allows us to independently vary the barriers in the $x$ and $y$ directions. Unless 
otherwise stated the parameters for our double-plaquette are: $\hbar\omega=1$ kHz,
$l_0=1.296$ $\mu$m, $d_w=6$ $\mu$m; (intraplaquette), $V_b=1.34$ kHz (intraplaquette)
$(\epsilon_b=0.5)$, $\sigma = 0.1833$ $\mu$m for all potential wells, and total-spin projection $S_z=0$. 
The $V_b$ between the plaquettes (interplaquette) varies depending on the interplaquette distance 
($\epsilon_b=0.5$ is kept constant). 

\subsection{Solution of the single particle Hamiltonian}

For rapid convergence of the full CI calculations, it is often beneficial to use a basis that already 
diagonalizes the single particle part of the Hamiltonian. Since the single-particle Hamiltonians used in this 
paper are separable in $x$ and $y$ we have to solve two one-dimensional Schr\"{o}dinger equations. A software 
package that does so, with high performance and high accuracy, is Chebfun \cite{driscoll_chebfun_2014}. It uses 
Chebychev polynomials as an efficient and highly accurate approximation of the true wave functions. From the 
documentation:

"The implementation of Chebfun is based on the mathematical fact that smooth functions can be
represented very efficiently by polynomial interpolation in Chebyshev points, or equivalently, thanks
to the Fast Fourier Transform, by expansions in Chebyshev polynomials"; and "Chebfun makes use of adaptive 
procedures that aim to find the right number of points automatically so as to represent each function to roughly 
machine precision, that is, about 15 digits of relative accuracy."

We have confirmed the accuracy of these solutions with both symbolic and numeric calculations in MATHEMATICA and 
the agreement was excellent at significantly higher performance.

\section{Analytical and numerical solutions of the Hubbard Hamiltonian}

To obtain numerical solutions to the Hubbard model we use the software package SNEG \cite{zitko_sneg_2011}, 
which is a "MATHEMATICA package for symbolic calculations with second-quantization-operator expressions". From 
the documentation: 

"SNEG library is a MATHEMATICA package that provides a framework for performing calculations using the operators 
of the second quantization with an emphasis on the anti-commuting fermionic operators in the context of 
solid-state and atomic physics. It consists of a collection of transformation rules that define the algebra of 
operators and a comprehensive library of utility functions."

\noindent For the case of a single square plaquette, the analytical results in \cite{schumann_2012} were also 
consulted.


\begin{thebibliography}{99}
\bibitem{joch12}
G. Z\"{u}rn, F. Serwane, T. Lompe, A. N. Wenz, M. G. Ries,
J. E. Bohn, and S. Jochim, 
Fermionization of two distinguishable fermions,
Phys. Rev. Lett. {\bf 108}, 075303 (2012).
\bibitem{joch15}
S. Murmann, A. Bergschneider, V. M. Klinkhamer, G. Z\"{u}rn, T. Lompe, and S. Jochim, 
Two fermions in a double well: exploring a fundamental building block of the Hubbard model,
Phys. Rev. Lett. {\bf 114} 080402 (2015).
\bibitem{joch15.2}
S. Murmann, F. Deuretzbacher, G. Z\"{u}rn, J. Bjerlin, S. M. Reimann, L. Santos, 
T. Lompe, and S. Jochim, 
Antiferromagnetic Heisenberg spin chain of a few cold atoms in a one-dimensional trap,
Phys. Rev. Lett. 115, 215301 (2015). 
\bibitem{kauf14}
A. M. Kaufman, B. J. Lester, C. M. Reynolds, M. L. Wall, M. Foss-Feig, K. R. A. Hazzard, A. M. Rey 
and C. A. Regal,
Two-particle quantum interference in tunnel-coupled optical tweezers,
Science {\bf 345}, 306 (2014).
\bibitem{kauf15}
A. M. Kaufman, B. J. Lester, M. Foss-Feig, M. L. Wall, A. M. Rey and C. A. Regal,
Entangling two transportable neutral atoms via local spin exchange,
Nature {\bf 527} 208 (2015).
\bibitem{hule15}
R. A. Hart, P. M. Duarte, T.-L. Yang, X. Liu, Th. Paiva, E. Khatami, R. T. Scalettar,
N. Trivedi, D. A. Huse, and R. G. Hulet,
Observation of antiferromagnetic correlations in the Hubbard model with ultracold atoms,
Nature {\bf 519}, 211 (2015).
\bibitem{bloc16}
M. Boll, T. A. Hilker, G. Salomon, A. Omran, J. Nespolo,
L. Pollet, I. Bloch, and C. Gross,
Spin- and density-resolved microscopy of antiferromagnetic correlations in Fermi-Hubbard chains, 
Science {\bf 353}, 1257 (2016).
\bibitem{grei16}
M. F. Parsons, A. Mazurenko, C. S. Chiu, G. Ji, D. Greif, and M. Greiner,
Site-resolved measurement of the spin-correlation function in the Fermi-Hubbard model,
Science {\bf 353}, 1253 (2016).
\bibitem{zwie16}
L. W. Cheuk, M. A. Nichols, K. R. Lawrence, M. Okan, H. Zhang, E. Khatami, N. Trivedi, T. Paiva, M. Rigol,
and M. W. Zwierlein,
Observation of spatial charge and spin correlations in the 2D Fermi-Hubbard model, 
Science {\bf 353}, 1260 (2016).
\bibitem{feyn82} 
R. P. Feynman,
Simulating physics with computers,  
Int. J. Theor. Phys. {\bf 21}, 467 (1982).
\bibitem{yann15}
B. B. Brandt, C. Yannouleas, and U. Landman,
Double-Well ultracold-fermions computational microscopy: Wave-function anatomy of attractive-pairing and
Wigner-molecule entanglement and natural orbitals,
Nano Lett. {\bf 15}, 7105 (2015).
\bibitem{poll15}
M. A. Garc\'{i}a-March, A. Yuste, B. Juli\'{a}-D\'{i}az, and A. Polls,
Mesoscopic superpositions of Tonks-Girardeau states and the Bose-Fermi mapping,
Phys. Rev. A {\bf 92}, 033621 (2015).
\bibitem{yann16}
C. Yannouleas, B. B. Brandt, and U. Landman,
Ultracold few fermionic atoms in needle-shaped double wells: spin chains and resonating spin clusters 
from microscopic Hamiltonians emulated via antiferromagnetic Heisenberg and $t$-$J$ models, 
New J. Phys. {\bf 18}, 073018 (2016).
\bibitem{yann07}
C. Yannouleas and U. Landman,
Symmetry breaking and quantum correlations in finite systems: studies of quantum dots 
and ultracold Bose gases and related nuclear and chemical methods,
Rep. Prog. Phys. {\bf 70}, 2067 (2007).
\bibitem{yann09}
Ying Li, C. Yannouleas, and U. Landman,
Artificial quantum-dot helium molecules:
Electronic spectra, spin structures, and Heisenberg clusters,
Phys. Rev. B {\bf 80}, 045326 (2009).
\bibitem{szabobook}
A. Szabo and N.S. Ostlund,
{\it Modern Quantum Chemistry\/} (McGraw-Hill, New York, 1989).
\bibitem{ande08}
P. W. Anderson, P. A. Lee, M. Randeria, T. M. Rice, N. Trivedi, and F. C. Zhang,
The physics behind high-temperature superconducting Cuprates: The ``Plain Vanilla'' version of RVB,
J. Phys. Cond. Mat. {\bf 16}, R755 (2004);
P. W. Anderson,
Who or what is RVB?,
Phys. Today {\bf 61}, 8 (2008).
\bibitem{fuku08}
M. Ogata and H. Fukuyama,
The $t$-$J$ model for the oxide high-T$_c$ superconductors,
Rep. Prog. Phys. {\bf 71}, 036501 (2008).
\bibitem{scal96}
D. J. Scalapino and S. A. Trugman,
Local antiferromagnetic correlations and $d_{x^2-y^2}$ pairing, 
Philos. Mag. B {\bf 74}, 607 (1996). 
\bibitem{tsai06}
W.-F. Tsai and S. A. Kivelson,
Superconductivity in inhomogeneous Hubbard models,
Phys. Rev. B {\bf 73}, 214510 (2006). 
\bibitem{tsai08}
W.-F. Tsai, 
Inhomogeneous Hubbard models
(Ph. D. Dissertation, University of California Los Angeles, 2008).
\bibitem{hofs02}
W. Hofstetter, J. I. Cirac, P. Zoller, E. Demler, and M. D. Lukin,
High-Temperature Superfluidity of Fermionic Atoms in Optical Lattices,
Phys. Rev. Lett. {\bf 89}, 220407 (2002).
\bibitem{zoll06}
S. Trebst, U. Schollw\"{o}ck, M. Troyer, and P. Zoller,
$d$-Wave resonating valence bond states of fermionic atoms in optical lattices,
Phys. Rev. Lett. {\bf 96}, 250402, (2006). 
\bibitem{rey09}
A. M. Rey, R. Sensarma, S. F\"{o}lling, M. Greiner, E. Demler, and M. D. Lukin,
Controlled preparation and detection of $d$-wave superfluidity in two-dimensional optical superlattices,
Eur. Phys. Lett. {\bf 87}, 60001 (2009).
\bibitem{pare08}
B. Paredes and I. Bloch,
Minimum instances of topological matter in an optical plaquette,
Phys. Rev. A {\bf 77}, 023603 (2008).
\bibitem{nasc12}
S. Nascimb\`{e}ne, Y.-A. Chen, M. Atala, M. Aidelsburger, S. Trotzky, B. Paredes, and I. Bloch,
Experimental realization of plaquette resonating valence-bond states with ultracold atoms in optical 
superlattices,
Phys. Rev. Lett. {\bf 108}, 205301 (2012).
\bibitem{fali69}
L. M. Falicov and R. A. Harris,
Two-Electron Homopolar Molecule: A Test for Spin-Density Waves 
and Charge-Density Waves,
J. Chem. Phys. {\bf 51}, 3153 (1969).
\bibitem{lueh15}
O. Dutta, M. Gajda, Ph. Hauke, M. Lewenstein, D.-S. L\"{u}hmann, B. A. Malomed, T. Sowi\'{n}ski 
and J. Zakrzewski,
Non-standard Hubbard models in optical lattices: a review,
Rep. Prog. Phys. {\bf 78}, 066001 (2015).
\bibitem{rice88}
F. C. Zhang and T. M. Rice,
Effective Hamiltonian for the superconducting Cu oxides,
Phys. Rev. B {\bf 37}, 3759(R) (1988).
\bibitem{note1}
The mass of $^6$Li is taken to be 10964.90$m_e$, with $m_e$ being the free-electron mass.
This gives $l_0=1.296$ $\mu$m for $\hbar \omega=1$ kHz. 
\bibitem{doga13}
R. A. Doganov, S. Klaiman, O. E. Alon, A. I. Streltsov, and L. S. Cederbaum,
Two trapped particles interacting by a finite-range two-body potential in two spatial dimensions,
Phys. Rev. A {\bf 87}, 033631 (2013).
\bibitem{yann99}
C. Yannouleas and U. Landman, 
Spontaneous symmetry breaking in single and molecular quantum dots,
Phys. Rev. Lett. {\bf 82}, 5325 (1999); Erratum {\bf 85}, 2220 (2000).
\bibitem{suppl}
See supplemental material (in Appendices D and E).
\bibitem{yann07.2}
L. O. Baksmaty, C. Yannouleas, and U. Landman,
Rapidly rotating boson molecules with long- or short-range repulsion:
An exact diagonalization study,
Phys. Rev. A {\bf 75}, 023620 (2007).
\bibitem{triv04}
A. Paramekanti, M. Randeria, and N. Trivedi,
High-T$_c$ superconductors: A variational theory of the superconducting state,
Phys. Rev. B {\bf 70}, 054504 (2004).
\bibitem{spal17}
J. Spa\l{}ek, M. Zegrodnik, and J. Kaczmarczyk,
Universal properties of high-temperature superconductors from real-space pairing: $t$-$J$-$U$ model and its 
quantitative comparison with experiment,
Phys. Rev. B {\bf 95}, 024506 (2017). 
\bibitem{schum02}
R. Schumann,
Thermodynamics of a 4-site Hubbard model by analytical diagonalization,
Ann. Phys. (Leipzig) {\bf 11}, 49 (2002).
\bibitem{lee06}
P. A. Lee, N. Nagaosa, and X. G. Wen,
Doping a Mott insulator: Physics of high-temperature superconductivity,
Rev. Mod. Phys. {\bf 78}, 17 (2006).
\bibitem{lee08}
P. A. Lee,   
From high temperature superconductivity to quantum spin liquid: 
progress in strong correlation physics,
Rep. Prog. Phys. {\bf 71}, 012501 (2008).
\bibitem{hubb78}
J. Hubbard, 
Generalized Wigner lattices in one dimension and some applications to tetracyanoquinodimethane 
(TCNQ) salts,
Phys. Rev. B {\bf 17}, 494 (1978).
\bibitem{baer09}
D. Baeriswyl D. Eichenberger, and M. Menteshashvili,
Variational ground states of the two-dimensional Hubbard model,
New J. of Phys. {\bf 11}, 075010 (2009).
\bibitem{yoko06}
H. Yokoyama, M. Ogata, and Y. Tanaka,
Mott transitions and $d$-wave superconductivity in half-filled-band Hubbard model on square lattice
with geometric frustration,
J. Phys. Soc. Japan {\bf 75}, 114706 (2006).
\bibitem{mezz16}
F. Mezzacapo, A. Angelone, and G. Pupillo,
Two holes in a two-dimensional quantum antiferromagnet: A variational study based on 
entangled-plaquette states,
Phys. Rev. B {\bf 94}, 155120 (2016).
\bibitem{rega16}
C. Regal, 
Bringing order to neutral atom arrays, 
Science {\bf 354}, 972 (2016).
\bibitem{barr16}
D. Barredo, S. de L\'{e}s\'{e}leuc, V. Lienhard, Th. Lahaye, and A. Browaeys,
An atom-by-atom assembler of defect-free arbitrary two-dimensional atomic arrays,
Science {\bf 354}, 1021 (2016).
\bibitem{endr16}
M. Endres, H. Bernien, A. Keesling, H. Levine, E. R. Anschuetz, 
A. Krajenbrink, C. Senko, V. Vuletic, M. Greiner, M. D. Lukin
Atom-by-atom assembly of defect-free one-dimensional cold atom arrays,
Science {\bf 354}, 1024 (2016). 
\bibitem{arpack}
R. B. Lehoucq, D. C. Sorensen, and C. Yang, 
{\it ARPACK USERS' GUIDE: Solution of large-scale eigenvalue
problems with implicitly restarted ARNOLDI methods\/}
(SIAM, Philadelphia, 1998).
\bibitem{lidis09}
Ying Li,
Confined quantum fermionic systems,
(Ph. D. Dissertation, Georgia Institute of Technology, 2009). 
\bibitem{paunczbook}
R. Pauncz, 
{\it Spin Eigenfunctions, Construction and Use\/}
(Plenum Press, New York, 1979).
\end{thebibliography}

\begin{thebibliography}{29}
\bibitem{li_artificial_2009}
Ying Li, C. Yannouleas, and U. Landman,
Artificial quantum-dot helium molecules:
Electronic spectra, spin structures, and Heisenberg clusters,
Phys. Rev. B {\bf 80}, 045326 (2009).
\bibitem{brandt_double-well_2015}
B. B. Brandt, C. Yannouleas, and U. Landman,
Double-Well ultracold-fermions computational microscopy: Wave-function anatomy of attractive-pairing and
Wigner-molecule entanglement and natural orbitals,
Nano Lett. {\bf 15}, 7105 (2015).
\bibitem{yannouleas_ultracold_2016}
C. Yannouleas, B. B. Brandt, and U. Landman,
Ultracold few fermionic atoms in needle-shaped double wells: spin chains and resonating spin clusters
from microscopic Hamiltonians emulated via antiferromagnetic Heisenberg and $t$-$J$ models,
New J. Phys. {\bf 18}, 073018 (2016).
\bibitem{driscoll_chebfun_2014}
T. A. Driscoll, N. Hale, and L. N. Trefethen,
{\it Chebfun Guide} (Pafnuty Publications).
\bibitem{zitko_sneg_2011}
R. \v{Z}itko,
SNEG -- MATHEMATICA package for symbolic calculations with second-quantizationoperator
expressions,
Comp. Phys. Commun. {\bf 182}, 2259 (2011).
\bibitem{schumann_2012}
Schumann, R., 
Thermodynamics of a 4-site Hubbard model by analytical diagonalization,
Ann. Phys., {\bf 11}: 49–88 (2002).
\end{thebibliography}
\end{document}